\documentclass{aa}
\usepackage{natbib}         
\usepackage{graphicx}                    
\usepackage{amssymb}                    
\usepackage[usenames]{color}                         
\usepackage{txfonts}

\newcommand{\BE}{\begin{equation}}
\newcommand{\EE}{\end{equation}}
\newcommand{\BA}{\begin{eqnarray}}
\newcommand{\EA}{\end{eqnarray}}
\newcommand{\fig}[1]{Fig.~\ref{fig_#1}}
\newcommand{\figs}[2]{Figs.~\ref{fig_#1} and \ref{fig_#2}}

\newcommand{\sect}[1]{Sect.~\ref{sec_#1}}

\newcommand{\eg}{e.g.}
\newcommand{\ie}{i.e.}

\newcommand{\cf}{cf.}

\newcommand{\degree}{\ensuremath{^\circ}}

\begin{document}

\title{Blowout Jets and Impulsive Eruptive Flare in a Bald-Patch Topology}

\titlerunning{Blowout jets and eruptive flare}

\author{R. Chandra\inst{1} \and C.H. Mandrini\inst{2,3} \and B. Schmieder\inst{4}
\and B. Joshi\inst{5} \and G.D. Cristiani\inst{2,3} \and H. Cremades\inst{6} \and E. Pariat\inst{4}
\and F.A. Nuevo\inst{2,3}  \and A.K. Srivastava\inst{7}
 \and W. Uddin\inst{8}  
            }
\offprints{R. Chandra}

\institute{
$^{1}$ Department of Physics, DSB Campus, Kumaun University, Nainital, India, 263 002 
\email{rchandra.ntl@gmail.com}\\
$^{2}$ Instituto de Astronom\'\i a y F\'\i sica del Espacio, UBA-CONICET, CC. 67, Suc. 28, 1428 Buenos Aires, Argentina \email{mandrini@iafe.uba.ar, gcristiani@iafe.uba.ar, federiconuevo@gmail.com}\\
$^{3}$ Departamento de F\'\i sica, Facultad de Ciencias Exactas y Naturales, Universidad de Buenos Aires, 1428 Buenos Aires, Argentina \\
$^{4}$ LESIA, Observatoire de Paris, PSL Research University, UMR 8109 (CNRS), Sorbonne Universit\'es, UPMC Univ. Paris 06, Univ. Paris Diderot, Sorbonne Paris 
Cit\'e, 5 place Jules Janssen, F-92195 Meudon, France \email{Brigitte.Schmieder@obspm.fr, Etienne.Pariat@obspm.fr}\\
$^{5}$ Udaipur Solar Observatory, Physical Research Laboratory, Udaipur, India -313 004\\
$^{6}$ Universidad Tecnol\'ogica Nacional, Facultad Regional Mendoza, CONICET, CEDS, Mendoza, Argentina \email{hebe.cremades@frm.utn.edu.ar}\\
$^{7}$ Department of Physics, Indian Institute of Technology (Banaras Hindu University), Varanasi, India- 221 005\\
$^{8}$ Aryabhatta Research Institute of Observational Sciences (ARIES) Nainital, India- 263 001\\
}

\date{Received ***; accepted ***}

   \abstract
{A subclass of broad EUV and X-ray jets, called blowout jets, have become a topic of research since they could be the link between standard collimated jets and CMEs.}
{Our aim is to understand the origin of a series of broad jets, some accompanied by flares
and associated with narrow and jet-like CMEs.} 
{We analyze observations of a series of recurrent broad jets observed in AR 10484 on 21\,--\,24 October 2003. In particular, one of them occurred simultaneously with an M2.4 flare on 23 October at 02:41 UT (SOLA2003-10-23). Both events were observed by ARIES H$\alpha$ Solar Tower-Telescope, TRACE, SOHO, and RHESSI instruments. The flare was very impulsive and followed by a narrow CME.
A local force-free model of AR 10484 is the basis to compute its topology. We find bald patches (BPs) at the flare site. This BP topology is present for at least two days before. Large-scale 
field lines, associated with the BPs, represent open loops. This is confirmed by a global PFSS model.
Following the brightest leading edge of the H$\alpha$ and EUV jet emission, we can temporarily associate it with a narrow CME.}
   {Considering their characteristics, 
the observed broad jets appear to be of the blowout class. 
As the most plausible scenario, we propose that magnetic reconnection could occur at the BP separatrices forced by the destabilization of a continuously reformed flux rope underlying them. 
The reconnection process could bring the cool flux-rope material into the reconnected open field lines driving the series of recurrent blowout jets and accompanying CMEs.}
{Based on a model of the coronal field, we compute the AR 10484 topology at the location where flaring and blowout jets occurred from 21 to 24 October 2003. This topology can consistently explain the origin of these events.  
 }

\keywords{Sun: magnetic fields, Sun: flares, Sun: jets,  Sun: coronal mass ejections (CMEs), Sun: X-rays, gamma rays}

\maketitle

\section{Introduction} 
\label{sec_introduction}

Plasma ejected from the Sun in transient phenomena is observed over a large spatial, temporal, and wavelength range. From full-fleshed bright coronal mass ejections (CMEs) going through faint CMEs and narrow CMEs (observed in white-light coronagraphs) to blowout and the more standard collimated jets seen in X-rays and EUV, there seems to exist a continuum of ejective events playing a key role in the dynamics of the corona and the solar wind.

\citet{Hundhausen1984} formulated the classic definition of CME as ``an observable change in coronal structure that (1) occurs on a timescale between a few minutes and several hours and (2) involves the appearance of a new, discrete, bright white-light feature in the coronagraph field of view'', which was later updated by \citet{Schwenn2006}. However, a wealth of morphologies complies with this definition, which inspired several attempts to categorize solar ejections, such as those by \citet{Howard1985} and \citet{Burkepile1993}, including classes such as loop, halo, fan, blob, jet,  spike, among others. With time, the concept of a regular CME has evolved to refer to a bubble-like eruption of significant angular width 
(\ie~$>$ 40$^0$), which may show the three-part structure of a bright leading edge, a dark void, and a bright core 
\citep[three-part CME; ][]{Illing1985}, or simply ,a bright front after which a clear cavity and/or a bright core cannot be discerned (loop-CME). Recently, \citet{Vourlidas2013} classified CMEs reported by the Coordinated Data Analysis Workshops (CDAW) CME Catalog \citep[http://cdaw.gsfc.nasa.gov/CME list;][]{Yashiro2004}, from the Large 
Angle and Spectroscopic Coronagraph (LASCO) instrument \citep{Brueckner1995} onboard Solar and Heliospheric Observatory (SOHO), into categories exhibiting a flux-rope structure, namely the three-part and the loop CME types, and categories without a clear flux rope structure, which included jets and outflows. Their jet category includes events with angular widths
$\leq$ 40$^0$, lacking a sharp front and detailed sub-structure or circular morphology, while their outflow category comprises wider events without a clear loop front or cavity that may contain filamentary material. \citet{Yashiro2003} investigated the properties of hundreds of narrow CMEs of the CDAW SOHO/LASCO Catalog, with narrow CMEs being defined as events with angular width $\leq$ 20$^0$. They recognized that many narrow CMEs may not be listed in the catalog because they are too faint and it is difficult to identify them. Faint CMEs of all widths may be attributed either to a scarcity of material associated with the event or to a significant propagation component away from the observer's plane of the sky (POS) \citep[\eg,][]{Cremades2015}.

Coronal jets were first observed in detail by the Soft X-ray Telescope (SXT) onboard the Japanese mission Yohkoh. Yohkoh/SXT data revealed the characteristics of large and energetic jets \citep{Shibata1992a,Shimojo1996,Shimojo1998,Shimojo2000}. Observations in X-rays and EUV show that they are collimated, beam-like structures, typically anchored in coronal bright points, and best seen inside polar coronal holes against the dark background. Polar coronal jets were studied by \citet{Wang1998} using data from LASCO and the Extreme Ultraviolet Imaging Telescope (EIT) onboard SOHO. Hinode has also provided data on polar jet parameters, such as lifetime, length, transverse width, and speed \citep{Savcheva2007,Moreno-Insertis2008,Filippov2009}. Typical jet durations are from tens of minutes to more than one hour \citep{Savcheva2007,Cirtain2007}, lengths are in the range  1--20 $\times$ 10$^4$ km, widths in the range 1--4 $\times$ 10$^4$ km, and speeds range from 150 km s$^{-1}$ to over 800 km s$^{-1}$ \citep{Shimojo1996,Cirtain2007}. \citet{Nistico2009} built the first catalog of polar jets observed by the Sun Earth Connection Coronal and Heliospheric Investigation (SECCHI) onboard the Solar Terrestrial Relations Observatory (STEREO), while \citet{Nistico2010} did the same for equatorial jets finding no difference between their typical characteristics.  
\citet{Moore2010} introduced the concept of blowout jet to distinguish a series of events among the just described jets, and named the remaining ones standard jets. We follow the nomenclature of these authors throughout this article. Blowout jets show an initial phase similar to that of standard jets starting with a brightening at its base and a narrow spire; this is followed by a violent flux rope eruption and the consequent broadening of the spire \citep[see also][]{Sterling2010,Liu2011,Madjarska2011,Moore2013}. For a review on observations, theory, and modeling of coronal jets see \citet{Raouafi2016}.        

Magnetic reconnection is the mechanism generally accepted for energy release during jets, flares, and several CME models \citep[see the review by][for CME initiation mechanisms]{Schmieder2015}. 
In what kind of magnetic configurations they occur and how is reconnection forced or driven to trigger these active events give raise to
different theoretical and numerical models. Models used to explain jets differ according to the flow pattern observed and/or imposed at the photosphere and their magnetic configuration. In these lines, one finds emergence-flux models \citep{Heyvaerts1977,Shibata1992b} and cancelling-flux models \citep[][see also \cite{Longcope1998}]{Priest1994}. To drive magnetic reconnection in numerical simulations, \citet{Moreno-Insertis2008}, \citet{Moreno-Insertis2013}, 
\citet{Torok2009}, \citet{Gontikakis2009}, \citet{Archontis2010}, and \citet{Fangfang2014}, among others, propose the presence of flux emergence; while, \citet{Pariat2009b,Pariat2010,Pariat2015} force the magnetic configuration imposing horizontal photospheric twisting motions. The used magnetic field topology includes, in all cases, a coronal magnetic null-point; the model analyzes the build up of currents in its fan-separatrix surface and the consequent energy release resulting in heating and jet ejection. A topology with a null point is supported by a large number of observations since jets were first identified \citep[see the review by][and references therein]{Raouafi2016}. However, other magnetic topologies computed from coronal extrapolation models have been found associated with jet observations. Quasi-separatix layers were identified by \citet{Mandrini1996} and \citet{Guo2013}, with no magnetic null point found in any of these studies. In \citet{Guo2013}, as in \citet{Schmieder2013}, magnetic dips and bald patches were found in the configurations but their relation to the observed jets did not play any role or was marginal. Bald patches and associated separatrices have so far never been included in jet numerical simulations.              

Jets, standard or blowout, have been observed in H$\alpha$ or EUV in association with white-light CMEs whose eruption can be traced back to the same location in an AR \citep{Liu2005,Liu2008,Liu2015,Li2015}. In particular, \citet{Yan2012} studied two successive flares of X-ray class M that occurred in AR 10484 on 22 October 2003.  Both were accompanied by very broad EUV jets. The one associated to the first M1.4 flare was the most clearly seen. According to the outflow velocity measurements for this first event, these authors concluded that they could trace the broad jets up to the corona and link them to two narrow CMEs.

In this article, we describe a series of recurrent plasma ejections seen in EUV, with one observed also in H$\alpha$, from 21 to 24 October at the border of the major eastern AR 10484 sunspot. Several of the ejections are broader and more extended than standard jets. Furthermore, most of them can be associated to jet-like or narrow CMEs. Because of all these characteristics, we consider them to be jets of the blowout class. All the ejections were accompanied by simultaneous brightenings at their base, which in a few cases developed in M-class flares, as the ones discussed by \citet{Yan2012} and the M2.4 flare on 23 October (SOL2003-10-23) analyzed in this article. All these flares originated from approximately the same location in the AR and have similar characteristics.

Our article is organized as follows. We first describe the different data sets used in our analysis in \sect{obs}. In \sect{recurrent-jets}, we list all the events we identify that have an eruptive nature as those studied by \citet{Yan2012} and the one discussed in detail in \sect{multi}. After modeling the AR magnetic field and computing its topology in \sect{model}, we draw a scenario that can explain the origin of the minor flare brightenings, the flare described in \sect{multi} as well as those analyzed by \citet{Yan2012}, and the series of blowout jets associated to all these events. Finally, in \sect{conclusions} we discuss our results and conclude.   

\section{Observational data sets}
\label{sec_obs}
The observational data used in this study come from several sources. To understand the magnetic complexity of the active region and its relation to flare activity, we use the SOHO/Michelson ~Doppler ~Imager data \citep[MDI; ][]{Scherrer1995} with a cadence of 96~min and pixel resolution of 1.98$^{\prime\prime}$. H$\alpha$ observations were carried out at the Aryabhatta Research Institute of Observational Sciences (ARIES) in Nainital, India, using the 15 cm f/15 coud\`e
solar tower-telescope equipped with an H$\alpha$ filter with a passband of 0.5 \AA. The image size is enlarged by a 
factor of two using a Barlow lens. The pixel resolution of the images is 1$^{\prime\prime}$. To understand the thermal and non-thermal nature of the studied flare, we reconstructed X-ray images from the Reuven~Ramaty~ High-Energy ~Solar ~Spectroscopic
~Imager \citep[RHESSI;][]{Lin2002} using the CLEAN technique. The evolution of the flare and other associated dynamic phenomena in EUV are studied using data from the Transition Region and Coronal Explorer data \citep[TRACE;][]{Handy1999}. The pixel resolution of the images is 0.5 $^{\prime\prime}$ and the temporal cadence can be as low as tens of seconds. 
TRACE observed in six EUV/UV bands from Fe IX/X (171 \AA) to C I/Fe II/UV continuum (1600 \AA) to white-light emission. 
We complement TRACE observations with the SOHO/EIT \citep{Delaboudiniere1995} data. EIT observed
the full Sun with a cadence of 12~min and a pixel resolution of 2.36$^{\prime\prime}$ in four spectral bands centered on
Fe IX/X (171 \AA), Fe XII (195 \AA), Fe XV(284 \AA) and He II (304 \AA). The CMEs and minor ejections associated to the observed EUV jets are identified using images provided by the SOHO/LASCO C2 coronagraph, which observes the white-light corona from about 2 to 6 R$_{Sun}$ \citep{Brueckner1995}.

\begin{table*}
\caption{Recurrent jets in AR 10484}
\label{recurrent}
\centering
\begin{tabular}{|ccl|}
\hline
\multicolumn{3}{|c|}{\bf 21 October 2003}\\
\hline
Start time (UT)&Max. Length (Mm)&\multicolumn{1}{c|}{LASCO C2 feature}\\
\hline
14:24:10&44&*Faint narrow CME (AW<10) in PA=105, 14:54 UT\\
15:12:11&55&$^{Y}$Narrow CME (AW=18) in PA=98, 16:06 UT\\
17:48:12&48&Corona contaminated by CME from AR10486\\
\hline
\multicolumn{3}{|c|}{\bf 22 October 2003}\\
\hline
Start time (UT)&Max. Length (Mm)&\multicolumn{1}{c|}{LASCO C2 feature}\\
\hline
00:48:11&59&Corona contaminated by CME from AR10486 at 00:54 UT\\
02:36:11&168&Corona contaminated by CME from AR10486 at 03:54 UT\\
05:24:11&59&Corona contaminated by previous CME\\
09:48:11&217&Jet-like CME (AW$<$5) in PA=80, 10:30 UT\\
11:00:11&180&*Faint narrow CME (AW=10) in PA=90, 11:54 UT\\
14:24:11&61&Not evident, likely because faint, narrow, and high angle\\
15:12:11&190&$^{Y,R}$Narrow CME (AW=15) in PA=90, 15:30 UT\\
16:00:10&137&$^{Y}$Narrow CME (AW=20) in PA=85, 16:30 UT\\
\hline
\multicolumn{3}{|c|}{\bf 23 October 2003}\\
\hline
Start time (UT)&Max. Length (Mm)&\multicolumn{1}{c|}{LASCO C2 feature}\\
\hline
01:25:55&121&*Faint jet-like CME (AW$<$10) in PA=92, 2:06 UT\\
02:36:11&217&$^{Y,R}$Narrow CME (AW=20) in PA=90, 3:06 UT\\
06:36:11&82&$^{Y}$Narrow CME (AW=15) in PA=90, 7:31 UT\\
10:14:14&167&Jet-like CME (AW=10) in PA=80, 11:30 UT\\
11:48:11&243&Not distinguishable from previous jet, likely high angle\\
14:25:56&64&Not distinguishable from pre-existing structures, high angle\\
18:00:11&130&Not distinguishable from pre-existing structures, high angle\\
20:12:11&58&Corona contaminated by CME from AR10486 at 20:06 UT\\
22:00:11&166&Corona contaminated by CME from east limb, north of AR10486\\
23:12:11&76&Narrow CME (AW$<$5) in PA=78, 00:06 UT\\
\hline
\multicolumn{3}{|c|}{\bf 24 October 2003}\\
\hline
Start time (UT)&Max. Length (Mm)&\multicolumn{1}{c|}{LASCO C2 feature}\\
\hline
00:24:11&89&Not distinguishable from pre-existing structures, high angle\\
06:00:11&91&Not distinguishable from pre-existing structures, high angle\\
08:24:11&150&*Narrow CME (AW=10) in PA=102, 08:54 UT\\
09:48:11&149&Faint CME (AW=27) in PA=260, 11:06 UT, high angle\\
20:24:11&47&Jet-like CME (AW$<$5) in PA=67, 21:30 UT\\
21:36:11&141&Jet-like CME (AW$<$10) in PA=68, 22:30 UT\\
22:36:11&191&Jet-like CME (AW$<$5) in PA=70, 23:30 UT\\
\hline
\end{tabular}
\tablefoot{An * in the third column indicates that the association with the corresponding jet is marginal.
A superscript `Y' and `R' indicates that the coronal counterpart was respectively reported to the CDAW SOHO/LASCO and the CACTus CME catalogs. The words `high angle' indicate a large angle between the observed structure and the POS. The hours indicate the first time of appearance in LASCO C2 field-of-view (FOV). The angular width (AW) and the position angle (PA) are expressed in degrees, with the PA measured counter-clockwise from solar north.}
\end{table*}

\begin{figure*}
\centering
\includegraphics [width=15.0cm]{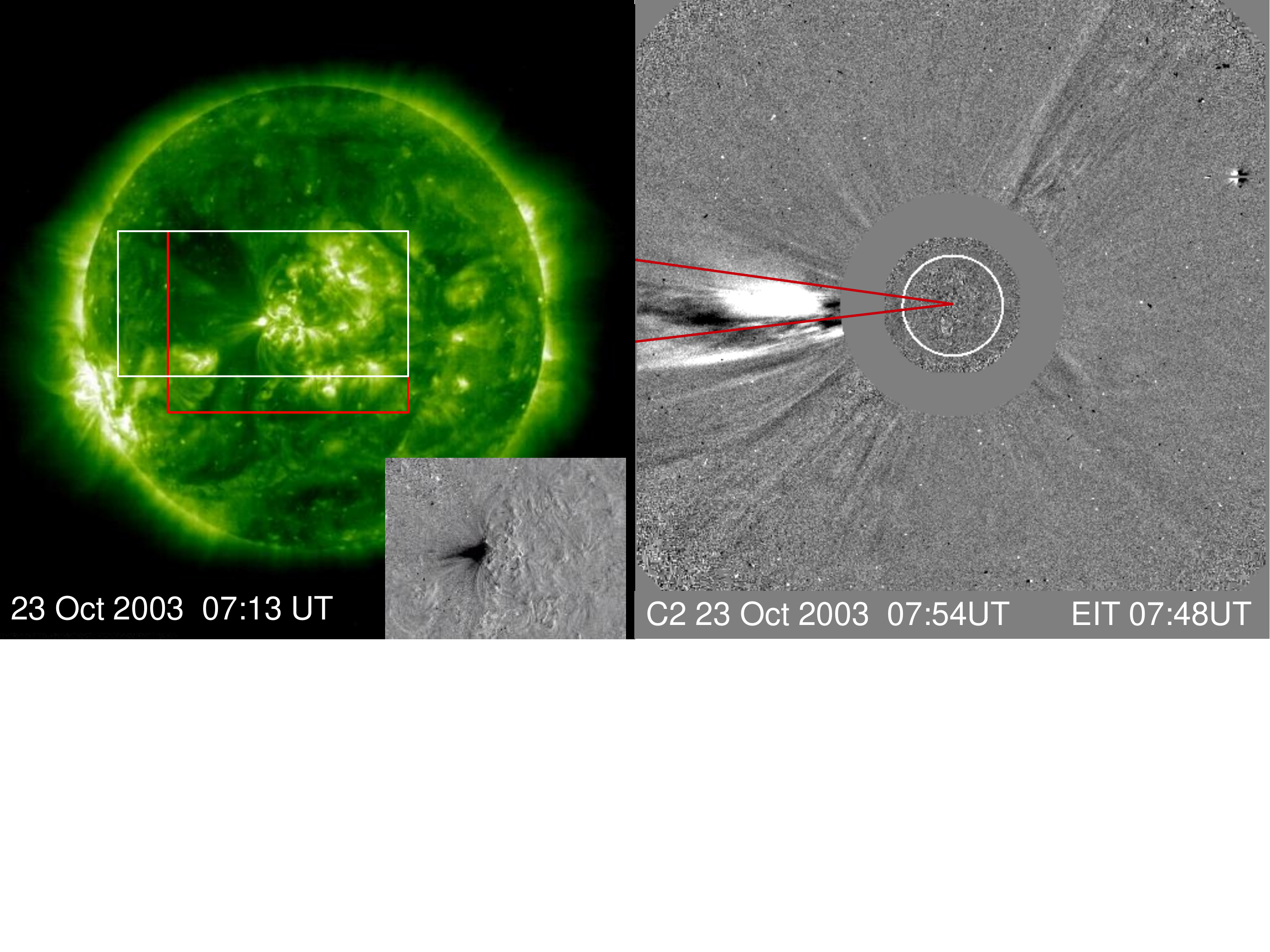}
\vspace{-3.5cm}
\caption{EIT image and LASCO C2 difference image of a jet and associated narrow CME.
(left panel) EIT full-disk image on 23 October 2003 at 07:13 UT showing a broad jet originating from the eastern border of AR 10484 at its maximum extension. The inset to its bottom right is the partial EIT difference image within the red rectangle in the full-disk image. (right panel) LASCO C2 difference image of the associated narrow CME. The two red continuous lines are drawn to indicate the CME angular width. Notice a very weak feature to the north of the CME bulk emission. This feature could appear due to projection effect because the CME  propagation direction has an important component towards the observer, as deduced from the source region location.
If we take it as the northern CME edge, then, 
the AW would be of $\approx$ 23 \degree, while the true width of the ejection is smaller. The broad feature to the south corresponds to a previous CME. The coronal evolution in the region indicated 
by the white rectangle is available as a movie in the online edition. These white and red rectangles partially overlap to the north and west.}
\label{fig_eit-jet}
\end{figure*}

\begin{figure*}[t]
\centering
\includegraphics [width=15.0cm]{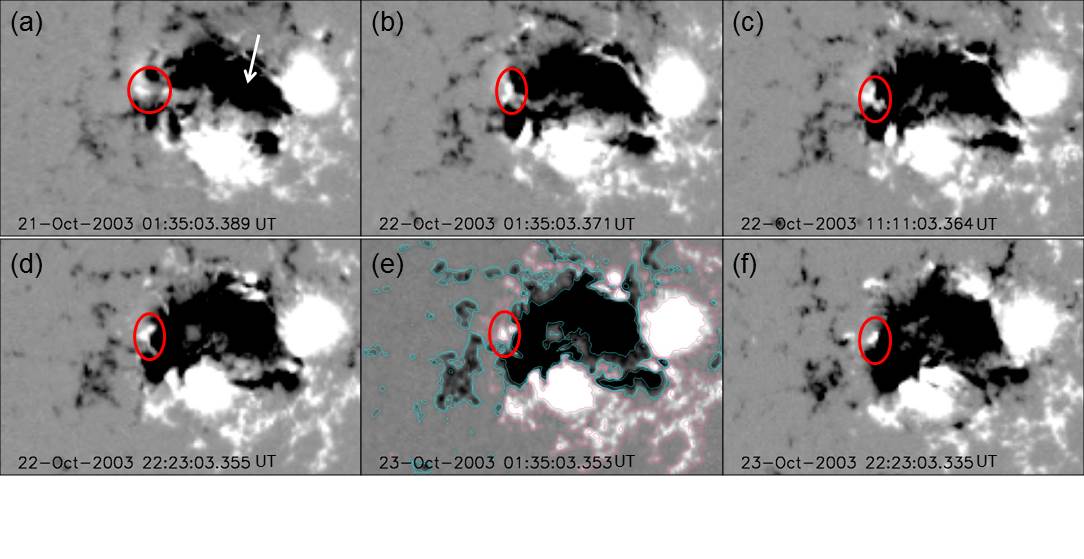}
\vspace{-1.cm}
\caption{Magnetic field evolution at the photospheric level. 
Panels (a)--(f) show the LOS magnetic field evolution as seen with MDI. We have marked with an arrow the location of a major negative sunspot around which a northern sunspot rotated from the north-east to the south-west. The ellipses indicate the approximate location of the flare and jet activity along those days. The images are derotated to Sun center and the size of the field of view (FOV)  is 290 per 200$^{\prime\prime}$. The images have been saturated below (above) -500 G (500 G). We have added three contours to panel (e) at $\pm 50$, $\pm 100$, and $\pm 500$ G in magenta (cyan) for positive (negative) values. }
\label{fig_mdi_evol}
\end{figure*}

\section{Recurrent broad jets in AR10484}
\label{sec_recurrent-jets}

The events discussed by \citet{Yan2012} and the ones analyzed in this article (see \sect{multi}) bear several characteristics in common. The flares (all M-class) started and evolved at approximately the same location in AR 10484 and were associated with broad jets that became CMEs, all of them within the categories named narrow and jet-like.
Motivated by these similarities, we investigate if similar phenomena occurred in AR 10484 between 21 to 24 October 2003. 
We search in EIT and TRACE EUV data and find that recurrent jets originated from the east of AR 10484 along these days. 
This tendency of H$\alpha$, EUV, and X-ray jets to appear emanating from  the same region recurrently has been observed in several examples \citep[e.g.][]{Schmieder1995,Asai2001,Wang2012,Zhang2012,Chandra2015,Panesar2016}. 
Our main aim in this section is to show the frequent occurrence of the ejection of material and to find evidences of that ejection in LASCO C2 images. We will not analyze all the events in detail in this section, but only those accompanying the M2.4 flare analyzed in \sect{multi}. 

All the jets we refer to in this section were broad and started with a small localized brightening at their base in EIT images. When TRACE images were available, this brightening appeared to have a loop-like shape. In some cases an X-ray class flare was reported in GOES data (see http://www.solarmonitor.org/ for the corresponding dates). In general and simultaneously with the EUV brightenings, a straight elongated mass expulsion was observed towards the east of AR 10484. This broad expulsion was observed sometimes in only one EIT image, either because its duration was less than 12 min or because the time span between consecutive EIT images was occasionally over 30 min. However, several jets lasted longer than 30 min. The EUV events can be seen in a movie in Fe XII 195 \AA~(eit-jets-21-24.mpeg), which is attached to this article. An example of one of these jets at its maximum extension, together with a LASCO C2 image showing its evidence in the upper corona, is depicted in \fig{eit-jet}.  
   
Table~\ref{recurrent} lists the broad jets observed from 21 October 2003 at 00:00 UT until the end of 24 October.  The list includes only the jets that are visible in 1024\,$\times$\,1024  EIT direct images and extend for at least 20 EIT pixels in those images. There are many more events originating from the east of the AR if one looks at EIT difference images. The first column in the table indicates the time at which we first detect the jet. These hours have an estimated error of 12 min, EIT 195 \AA~temporal cadence. 
Several of the broad jets, as the ones studied by \citet{Yan2012} and in this article, were also observed by TRACE with much higher temporal cadence but, since TRACE observed AR 10484 sporadically, we prefer to refer to EIT data for coherence in the table. The second column in Table~\ref{recurrent} shows the length of the jets measured in projection on the solar disk from EIT direct images. The lengths of the jets, reported at their maximum extensions, are measured from the brightenings observed at their bases to the brightest feature in their easternmost borders. These lengths have been calculated using standard SolarSoft routines and range from 44 to 243 Mm. 

Finaly, the third column in Table~\ref{recurrent} indicates if evidences of the ejected material is found in LASCO C2 images. We include a short description of what we observe analyzing LASCO C2 difference images, in combination with EIT difference images. 
We have consulted catalogs and data bases as an alternative to identify the white-light counterparts of AR 10484 broad jets. In particular, we looked at the CDAW SOHO/LASCO CME Catalog and the Computer Aided CME Tracking Catalog \citep[CACTus; http://sidc.oma.be/cactus/;][]{Robbrecht2004}. These catalogs rely on different techniques to recognize CME events, each of them with inherent limitations. The events we are looking for have a high chance to be small and narrow; therefore, most of the ones in Table~\ref{recurrent} are identified by us and do not appear in the catalogs. The few coronal counterparts reported by these catalogs have been indicated by a superscript `Y' and `R', respectively corresponding to the CDAW SOHO/LASCO and the CACTus CME catalogs. For the sake of consistency, we have measured ourselves the projected angular width (AW) and the position angle (PA) for all the LASCO C2 features reported in Table~\ref{recurrent}. 

Following the definition in \citet{Yashiro2003}, all except one of the coronal features are narrow CMEs according to their apparent AWs. However, they could be even narrower, because in correspondence to the location of the AR on the disk, the propagation direction of these events could have an important component towards SOHO, which produces a broader projected AW in the coronagraph images. Still, as hinted by the broad jets seen in projection on the solar disk in EIT images, the propagation direction of these events must have a significant component towards the east. We have indicated by `high angle' those white-light counterparts that presumably propagated at a larger angle with respect to the POS. Also in Table~\ref{recurrent}, `jet-like CME' indicates that the observed structure was long and very collimated, with hardly any evident structure, while a narrow CME was seen as being structured and in general having larger AWs than jet-like CMEs. The position angles (PAs) reported in Table~\ref{recurrent} lie within $\pm$ 5 \degree. 

We do not attempt to study in detail the temporal correlation between all the EUV broad jets and the LASCO C2 features listed in Table~\ref{recurrent}, instead we do this in \sect{multi} for the M2.4 flare and its associated jet on 23 October. We just aim to show that a white-light counterpart can be found for several of the broad jets observed from 21 to 24 October 2003 originating from the eastern side of AR 10484. By 25 October 2003, the broad-jet recurrency associated with white-light LASCO features is no longer evident \citep[see][]{Uddin2012}. 
          
\section{Multiwavelength view of the M2.4 flare and jet in AR 10484}
\label{sec_multi}

\subsection{Magnetic field evolution from 21 to 24 October 2003}
\label{sec_mag}

AR 10484 was a very complex region that appeared on the eastern solar limb on 18 October 2003 and could be followed until it rotated 
out to the far side of the Sun on 31 October. 
By 23 October it exhibited a $\beta\gamma\delta$ configuration (see http://www.solarmonitor.org/) and was formed by five major sunspots and several minor ones. It was a highly flare-productive AR during its disk transit whose events amounted to 29 C-class, 16 M-class, and 2 X-class flares \citep[see \eg,][for the analysis of some of these events]{Li2005,Zharkova2006,Yan2012,Uddin2012}. 

In \fig{mdi_evol} we show the evolution of the line of sight (LOS) magnetic field as seen with MDI from 21 October until late 23 October.   
This evolution was analyzed in detail in \citet{Yan2012}, together with the changes seen in TRACE white-light images. These authors observed the rotation towards the south-west of a large and a small negative sunspots located to the north-east of the AR  for around 48 hr in the counter-clockwise direction. The large sunspot rotated around the major northern negative spot (see their Fig. 1), which is pointed with a white arrow in \fig{mdi_evol}a. The rotation is also evident when one compares the different panels in \fig{mdi_evol}. As a result of this sustained rotation, the positive magnetic polarities located between the rotating spot and a southern and smaller negative one, as also other positive polarities to the east, were continuously cancelled and finally they practically disappeared, remaining visible only a small positive patch in \fig{mdi_evol}f.  A particular example of this cancellation and approach between positive and negative polarites can be seen in Fig. 2 of \citet{Yan2012}. \citet{Zhang2003} described the magnetic field configuration of AR 10484 using the Huairou Multi-Channel Solar Telescope. Their Fig. 1 clearly shows the counter-clockwise rotation of the horizontal component of the vector magnetic field on 21 and 22 October 2003 in a limited region of the AR, between the two major positive spots and including part of the rotating sunspot. These observations imply that the shear at the location of the rotating spot and surrounding field was high and was constantly being built up by the spot motion. 

In our view, the constant flux cancellation could be at the origin of the flaring and broad jet expulsions along this time period. We have added a red ellipse at approximately the location of EIT and TRACE brightenings and jets in panels a to f of \fig{mdi_evol}. The exact brightening location and jet origin may change within the ellipses considering the dynamic evolution of the magnetic field (see in the next sections the detailed analysis of the flare and jet at 02:41 UT on 23 October). Taking into account the numbers of events 
listed in Table \ref{recurrent}, the importance of the flares observed in AR10484 \citep[see][, and the flare analyzed in this article]{Yan2012}, and the MDI evolution in \fig{mdi_evol}, we conjecture that when magnetic flux cancellation was higher the activity with different brightness intensites and jets was more frequent and decreased as the positive flux mostly disappeared. 

In the aftermath of the broad jets and flares discussed in this article, \ie~on 25 October, H$\alpha$ and EUV activity in the form of minor flares and surges, continued to be observed to the east of the large major negative spot and to the north of the region within the red ellipse in \fig{mdi_evol}.
These phenomena were discussed by \citet{Uddin2012}, who suggested that magnetic reconnection between small dispersed polarities during the sunspot decay and prexisting fields could result in these active events. In these examples, the plasma is not injected into open field lines but into large closed loops that connect to the distant AR 10486 to the south-east of AR 10484 \citep[see Fig. 6 in][]{Uddin2012}.

\subsection{The flare and jet in X-rays and UV continuum}
\label{sec_x-ray}

\subsubsection{The temporal and spatial evolution}
\label{sec_ts-x-ray}

\begin{figure}
\centering
\includegraphics [width=8.cm]{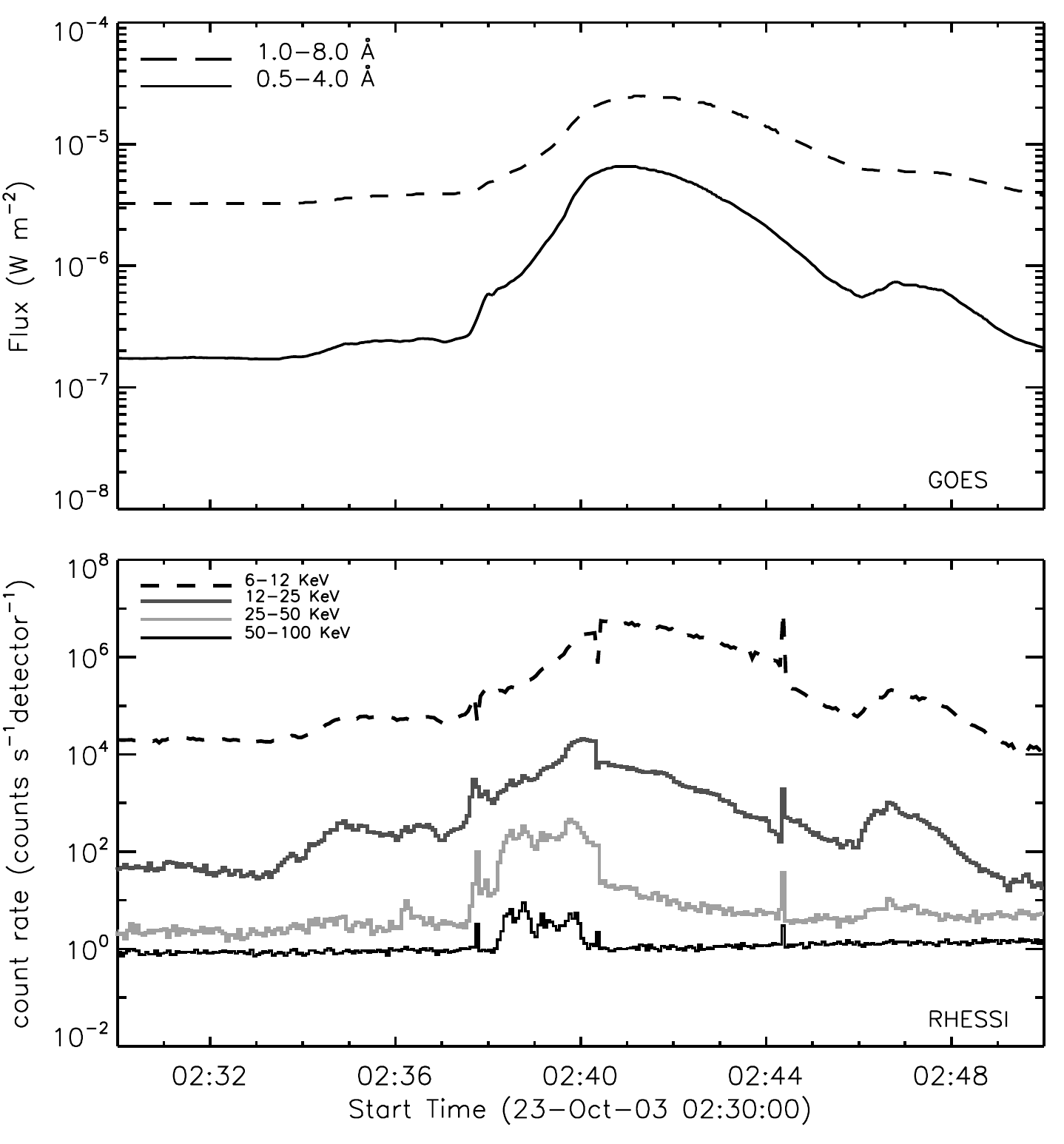}
\caption{GOES and RHESSI light curves of the M2.4 flare. RHESSI count rates are averaged over every 4 sec. To allow for the comparison of these curves, the RHESSI count rates are scaled by factors of 1, 1/5, 1/10, and 1/60 in the energy bands 
6--12, 12--25, 25--50, and 50--100 keV, respectively.}
\label{fig_x-ray_lc}
\end{figure}

The temporal evolution of the X-ray emision of the M2.4 X-class flare on 23 October is shown in \fig{x-ray_lc} 
using GOES (top) and RHESSI (bottom) data. GOES measurements in 
the 0.5--4.0~{\AA} channel show a small flux enhancement at $\sim$02:34 UT, 
indicating the onset of minor flaring in the AR. A small bump in GOES profiles at $\sim$02:38 UT, \ie~before the
maximum, is noteworthy. This is followed by a significant increase in the X-ray flux 
in GOES, as well as in RHESSI light curves. The GOES profiles peak around 02:41 UT. 

\begin{figure*}
\centering
\includegraphics [width=15.cm]{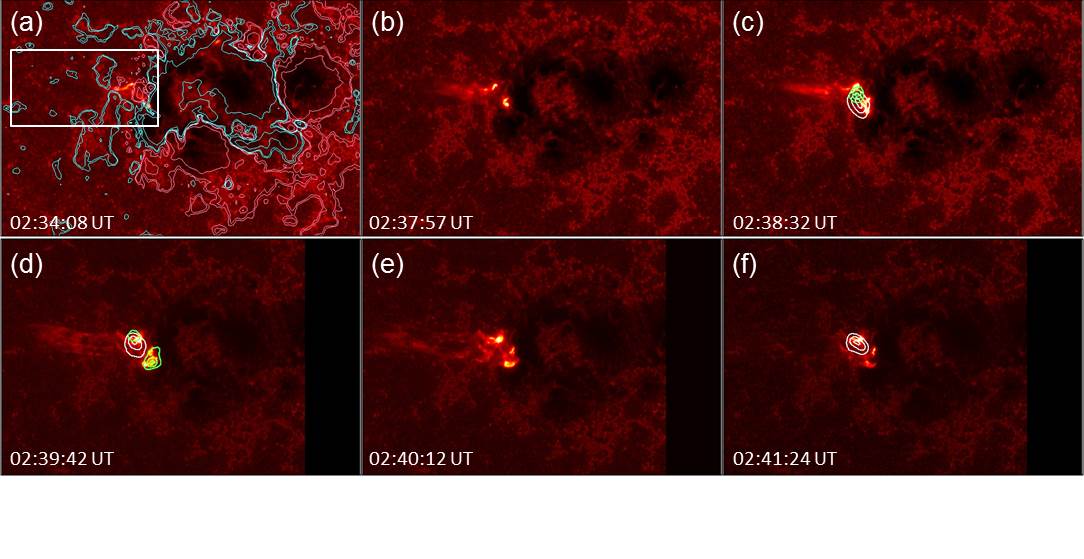}
\vspace{-1.cm}
\caption{Sequence of TRACE 1700~{\AA} images overlaid by co-temporal RHESSI X-ray contours. The yellow contours
correspond to the 10--15 keV range, while the light green contours to the 50--100 keV one. 
The integration time for each RHESSI image is 1 min. The contour levels are 40\%, 70\%, and 95\% of the peak flux in each image.
We have overlaid the same isocountors of the MDI magnetogram at 01:35 UT, shown in \fig{mdi_evol}e, to panel (a) in continuous magenta (cyan) line for positive (negative) field.
The field of view is the same as the one shown in \fig{mdi_evol}e and has been kept as such even when TRACE FOV size changed (see the black band to the west of panels (d), (e), and (f) and explanations in the text). The white rectangle in panel (a) indicates the FOV shown in \fig{bp-location}c, \fig{BP1-lines}a, and \fig{bp-lines-recon}a.}
\label{fig_trace_rhessi}
\end{figure*}

RHESSI \citep{Lin2002} observed this event from beginning to end. We analyze RHESSI  emission  in four energy bands,
namely, 6\,--\,12, 12\,--\,25, 25\,--\,50, and 50\,--\,100 keV. RHESSI light curves, shown in \fig{x-ray_lc} (bottom panel), are constructed
taking the average count rates over the front detectors 1, 3--6, 8, and 9 in each energy band. The low energy RHESSI light 
curves show an evolution similar to that of GOES. At higher energies, the light curves present an impulsive behavior. The flare 
produced significant hard X-ray (HXR) emission with energies up to $\approx$50--100 keV; this is quite interesting considering 
that the flare was quite compact (see \fig{trace_rhessi}). The early subtle rise observed between $\sim$02:34 and 02:38 UT in 
the 0.5--4.0~{\AA} 
GOES light curves is also quite prominent in the 12--25 keV X-ray energy band.

\fig{trace_rhessi} shows RHESSI contours in 10--15 and 50--100 keV overlaying TRACE images in 1700~\AA.
The RHESSI images are reconstructed using the CLEAN algorithm with the 
natural weighting scheme and the front detector segments 3 to 8 (excluding 7) \citep{Hurford2002}; this implies that RHESSI's spatial esolution is at most $\approx$ 7$^{\prime\prime}$.
The low energy emission  (\ie~10--15 keV) originates from a single compact source from the flare very 
beginning ($\sim$02:34 UT) until $\sim$02:48 UT. We have further
reconstructed images in other energy bands (not shown here), namely, 6--12, 15--25 keV and
found that the extension of the sources, as well as their spatial locations, at these energies are 
similar to the 10--15 keV source. 

A careful examination of series of TRACE 1700 {\AA} images shows that the earliest signature 
of flare activity appeared at $\sim$02:30 UT in the form of a localized brightening with a short
elongation towards the east of the AR. The TRACE FOV, with a size of 768 $\times$ 768 pixels, covered the full AR 10484 
until $\sim$2:39 UT, after that hour it reduced to 512 $\times$ 512 pixels leaving the west side of the AR 
without coverage.  At $\sim$02:35 UT, 
the intensity of the localized brightening, then seen as two separate kernels, increased suddenly and a 
jet-like ejection became very prominent. This phase is clearly reflected in the GOES X-ray profile 
as its initial bump. 

At $\sim$02:38 UT, the flaring region developed rapidly. The southern brightening 
became structured and splitted in two. With the continuous intensity increase, three bright kernels 
became evident. Between $\sim$02:38 and 02:40 UT, the HXR light curves show two impulsive peaks. 
Two HXR sources were observed for a brief period during this interval of time in the 50--100 keV energy band, 
these were approximately cospatial with the UV flare kernels (\fig{trace_rhessi}d).  
Although the flaring region is very compact, the low energy emission (10--15 keV)
at this time originated from a zone between the two HXR sources. We interpret that
the HXR sources, observed in the 50--100 keV energy band, mark the footpoints (FPs) of a closed loop-system 
at approximately the base of the jet in the compact flaring region. The HXR emission from the FPs is
traditionally viewed in terms of the thick-target bremsstrahlung process \citep{Brown1971} in which
X-ray production takes place when high-energy electrons, accelerated at the reconnection site, 
come along the guiding magnetic field lines and penetrate the denser transition region and chromospheric layers 
\citep{Kontar2010}. In view of this ``standard" picture, it is very likely that the HXR emission at lower X-ray energies 
(\eg, 10--15 keV) originated from the top of the flaring region. When examining the polarity of the field at the place 
of the high-energy HXR sources, we find that the sources are mostly located in regions of opposite polarity field (compare panels (a) and (d) in Fig. 4 and see panel (c) in Fig. 10).

\subsubsection{The spectral evolution}
\label{sec_spectra-x-ray}

We study the evolution of RHESSI X-ray spectra during the impulsive phase of the flare (\ie, between 02:37:00 UT and 
02:41:00 UT). For this analysis, we first generate a RHESSI spectrogram with an energy binning of $\frac{1}{3}$ keV from 6--15 
keV and 1 keV from 15--100 keV. We use only the front segments of the detectors and exclude detectors 2 and 7, which have 
lower energy resolution and high threshold energies, respectively. The spectra are deconvolved with the full detector response 
matrix \citep[\ie~off-diagonal elements are included,][]{Smith2002}. Spectral fits are obtained using a forward-fitting method 
implemented in the Object Spectral Executive (OSPEX) code. OSPEX allows the user to choose a model photon spectrum, which is convolved with the instrument response matrix and then fitted to the observed count spectrum. The best-fit parameters are obtained 
as output. We use the bremsstrahlung and line spectrum of an isothermal plasma and a power-law function with a turnover 
at low X-ray energies. 
The negative power-law index below the low energy turnover is fixed to 1.5. In this manner, there are five free parameters in 
the model: the temperature (T) and emission measure (EM) for the thermal component, the power-law index ($\gamma$), the normalization of the power-law, and the low energy turnover for the non-thermal component. From these fits we can derive the T and EM of the hot flaring plasma, as well as the power law index for the non-thermal component.

\begin{figure*}
\centering
\includegraphics [width=15cm]{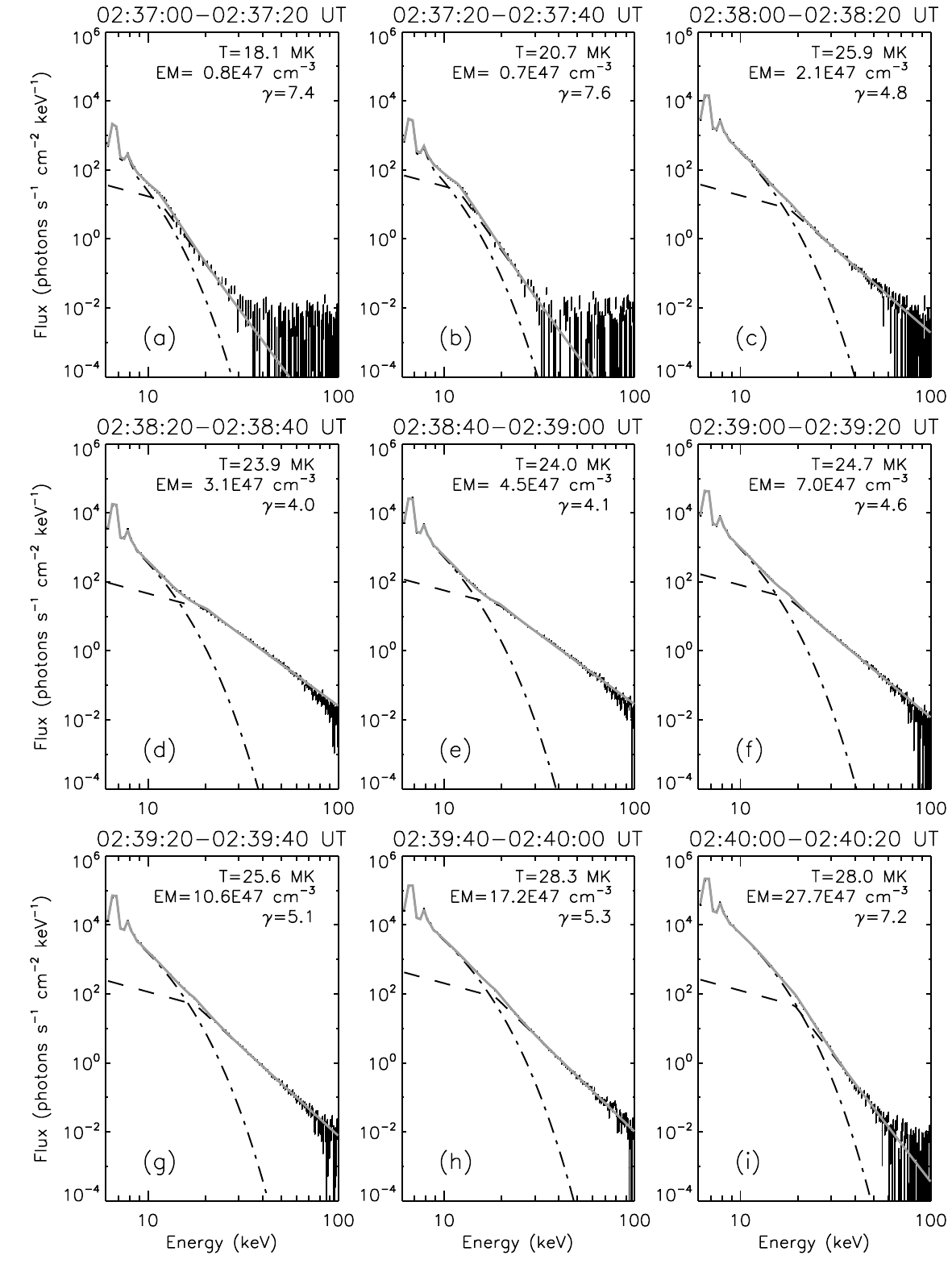}
\caption{ Hard X-ray spectra. 
RHESSI X-ray spectra derived during nine selected time intervals during the impulsive phase of the HXR emission together with 
the applied fits. The spectra are fitted with an isothermal model (dashed-dotted line) and a functional power-law with a turnover at low energies (dashed line). The gray solid line indicates the sum of the two components. The spectrum during the early impulsive phase (panels a and b) was fitted in the energy range 6--40 keV, while the other spectra (panels c--i) are fitted in the range 6--100 keV.}
\label{fig_rhessi_spec}
\end{figure*}

In \fig{rhessi_spec} we show the spatially-integrated and background-subtracted  RHESSI spectra derived during nine time intervals
of the impulsive phase of the flare together with the applied spectral fits. Each spectrum is accumulated over 20 sec. The RHESSI
X-ray time profiles reveal that the early impulsive phase emission ($\sim$02:37--02:38 UT) originates only at low X-ray energies
(\ie~$\approx$6--40 keV, \cf~\fig{x-ray_lc}). We find a rapid increase in the X-ray flux in HXR energies up to 50--100 keV after
$\sim$02:38 UT. Therefore, we have restricted the spectral fitting during the early phase (between 02:37 and 02:38 UT) to
the energy range 6--40 keV (\cf~\fig{rhessi_spec}a--b). The spectra derived during the main impulsive phase (between
02:38 and 02:41 UT) are fitted in the range 6--100 keV.

The spectra during the early phase (02:37:00--02:38:00 UT) show lower values of T and EM, indicating that the 
X-ray emission comes from localized regions. The TRACE UV images (see \fig{trace_rhessi}) clearly show 
localized brightenings in this interval.  
At X-ray energies $>$10 keV, the spectrum shows a steep power law with a photon spectral index of $\approx$7. The flare emission undergoes a very fast spectral evolution with the onset of high energy emission at $\sim$02:38 UT. The HXR emission is the  
hardest between 02:38:20 and 02:39:00 UT with a photon spectral index $\gamma$ of $\approx$4 (\cf~\fig{rhessi_spec}(d) 
and e). It is important to note that there is a decrease in temperature during this interval, indicating the dominance of 
the non-thermal characteristics of the emission. The temperature, as well as the emission measure, increase continuoulsy in the later stages while the slopes of HXR spectra steepen. This pattern in the evolution of HXR spectra, often called the soft-hard-soft 
spectral behavior, suggests that the phase of the hardest X-ray spectra is intimately associated with electron acceleration events
\citep{Grigis2004,Joshi2011,Joshi2012}.
The phase of the hardest X-ray emission is characterized by rapid changes in the flaring region in the UV images 
(\fig{trace_rhessi}) with the evolution of flare kernels. 
The HXR footpoints are briefly visible during this interval only.

\subsection{The flare and jet in H$\alpha$}
\label{sec_halpha}

\begin{figure*}
\centering
\includegraphics [width=15.0cm]{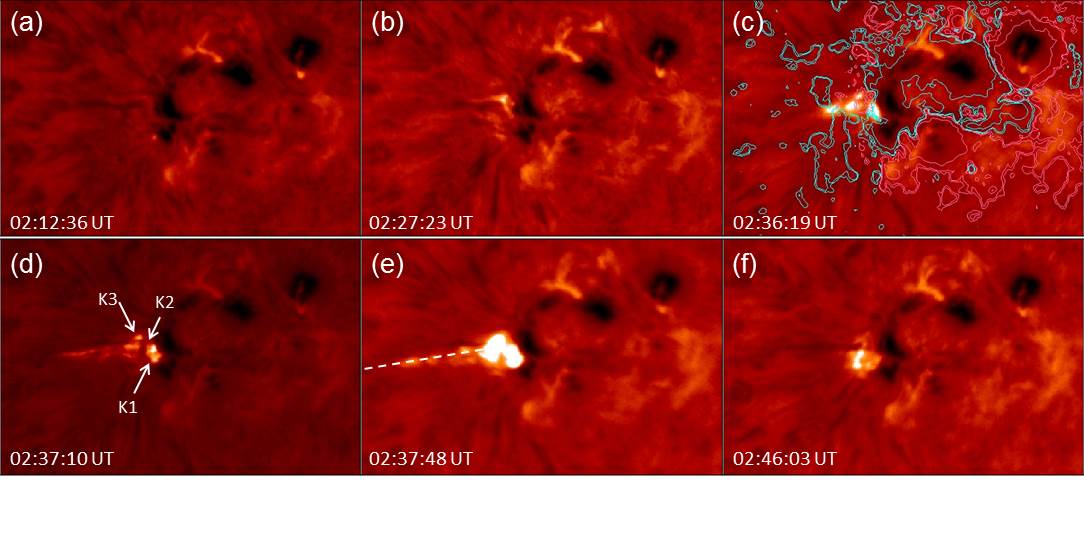}
\vspace{-1.cm}
\caption{Evolution of the flare in H$\alpha$. Three magnetic field contours ($\pm$ 50, $\pm$ 100, and $\pm$ 500 G) of the closest in time MDI map have been added to panel (c) in continuous magenta (cyan) line for positive (negative) field. The three brightest kernels have been labelled as K1, K2, and K3 in panel (d). All images have a FOV of 290 per 200$^{\prime\prime}$, similar to that of \fig{mdi_evol} and \fig{trace_rhessi}. The white dashed line in panel (e) indicates the direction along which the height-time measurements in \fig{slice} are performed.}
\label{fig_halpha}
\end{figure*}

The H$\alpha$ flare evolution is shown in \fig{halpha}.
In this wavelength, the flare initiated around 02:30 UT, peaked around 02:37 UT and
decayed around 02:50 UT. Similar to what happened in X-rays, the flare was very impulsive. It 
started with the ejection of dark, cool material towards the east. This was later followed 
by a bright, long, and broad jet as the material heated up when moving into the corona.  
The jet continued as the flare evolved and was no longer visible at 02:39 UT; however, 
the flare brightenings were visible up to 02:50 UT. The broad jet reached its maximum 
length during the flare peak. At the footpoints of the jet, we observe three compact H$\alpha$ kernels, called K1, K2, and K3 in \fig{halpha}d. Kernels K1 and K2 are located to the south-west and K3 to the north-east
of what seems to be the base of the jet, respectively. Initially kernels K1 and K2
appear as a single bright patch and later they separate. We have added three MDI contours to
\fig{halpha}c to show the flare brigthening locations relative to the magnetic field.

\subsection{The jet and its extension in a narrow CME} 
\label{sec_jet-lasco}

\begin{figure}
\centering
\includegraphics[width=8.0cm]{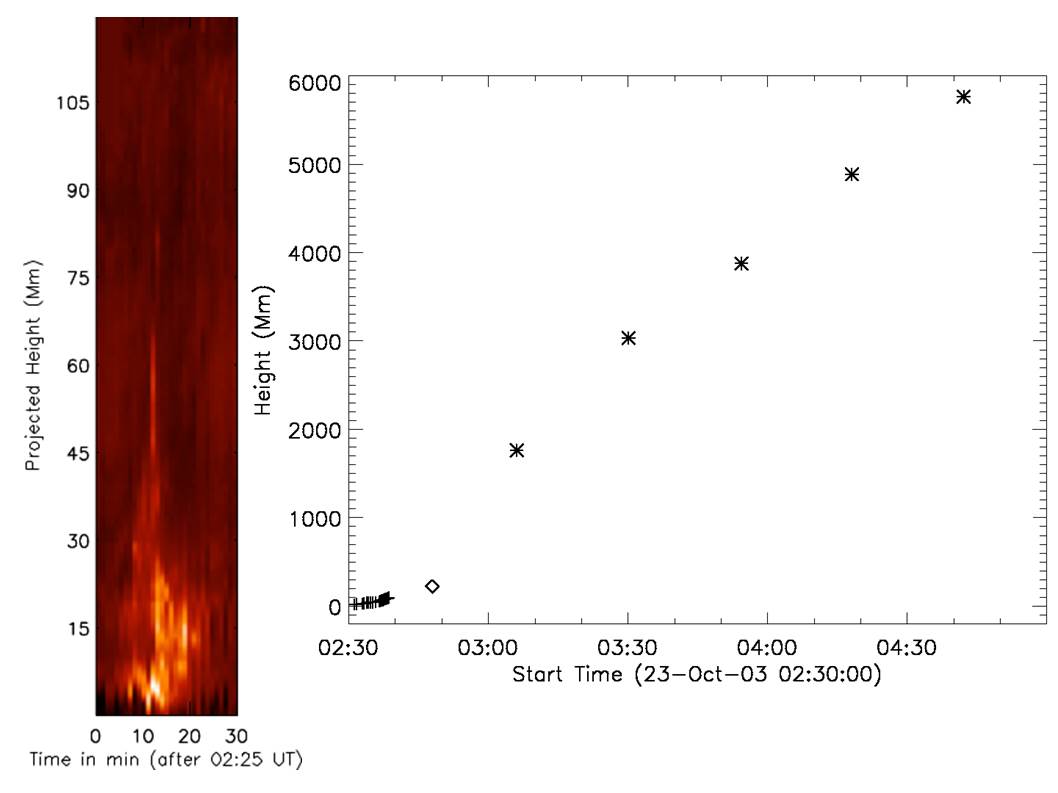}
\caption{ Projected height vs. time plots showing the link between the jet and the narrow CME. 
The left panel shows the time-slice diagram built along the white dashed line shown on the H$\alpha$ image at 02:37:48 UT (\fig{halpha}e). The right panel is a height-time plot combining the H$\alpha$ measurements of the jet length projected on the POS (vertical crosses), a similar measurement done on the EIT image at 02:48 UT (diamond), and those of the associated CME using data from the CDAW SOHO/LASCO Catalog (asterisks).}
\label{fig_slice}
\end{figure}

To study the kinematics of the jet, we create a time-slice diagram using the H$\alpha$ observations. 
The result of this analysis is presented in the left panel of \fig{slice}. The position of the slice along which we measure the distance of the brightest broad jet feature in function of time is shown by a white dashed line in panel (e) of  \fig{halpha}. Based on this analysis, we compute the projected velocity of the jet. This velocity is $\approx$ 220 km s$^{-1}$ and the maximum projected height achieved by the H$\alpha$ jet is 80 Mm with respect to its bright base (see also the height-time plot in the right panel of \fig{slice}).
Notice that we estimate a projected length of $\approx$ 215 Mm in the direct 195 \AA~ EIT images (see Table~\ref{recurrent}), which is consistent with the fact that the hot material is better seen in this wavelength range at greater coronal heights. As mentioned in \sect{recurrent-jets} the later estimation is done by measuring the distance of the brightest EUV feature in the jet with respect to its base at the time we observe the jet achieved its maximum length. Taking into account the poor temporal resolution of EIT images, we have not estimated a velocity from the EUV images. 

As shown in Table~\ref{recurrent}, the flare studied here can be associated with the narrow CME first seen in LASCO C2 at 03:06 UT for which we measure an AW of 20\degree. According to the CDAW SOHO/LASCO Catalog, the CME speed was 656 km s$^{-1}$. If we do a backward extrapolation of the LASCO Catalog's height-time plot, which exhibits an almost linear profile, we find that the approximate start time of the CME on the solar surface was 02:34 UT, which is consistent with the initiation time of the broad jet from the flaring region. 

\begin{figure*}
\centering
\includegraphics [width=15cm]{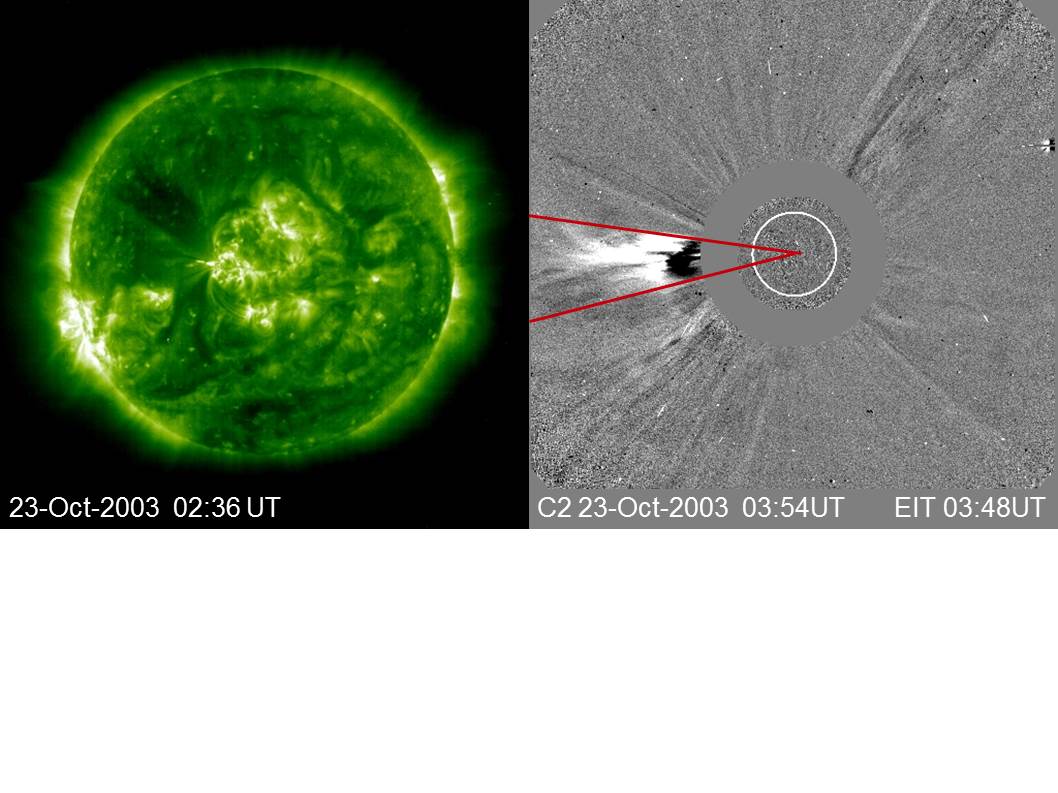}
\vspace{-3.cm}
\caption{ EIT image of the jet and LASCO C2 difference image of the associated narrow CME. (left panel) EIT full-disk image on 23 October 2003 at 02:38 UT showing the broad jet originating from the eastern border of AR 10484 during the M2.4 flare. (right panel) LASCO C2 difference image of the associated narrow CME. The two red solid lines are drawn to indicate the CME AW.}
\label{fig_eit_jet2}
\end{figure*}

In the right panel of \fig{slice}, we have added the EIT measurement of the projected jet height and of the associated CME, in the C2 FOV according to CDAW catalog, to the height-time plot of the H$\alpha$ jet. If we fit linearly the EUV measurement and LASCO C2 values, we find a velocity of $\approx$ 770 km s$^{-1}$ with a linear correlation coefficient of 0.99. This value is $\approx$ 17\%~larger than that of LASCO C2, which is reasonable considering the different kinds of measurements we are comparing.
However, when adding the H$\alpha$ measurement into the picture, we find that the ejected material first undergoes a phase of gradual acceleration, to reach later an almost constant speed. \fig{eit_jet2} illustrates the jet as observed with EIT at 02:36 UT and the associated narrow CME.  

\begin{figure*}
\centering
\includegraphics [width=15cm]{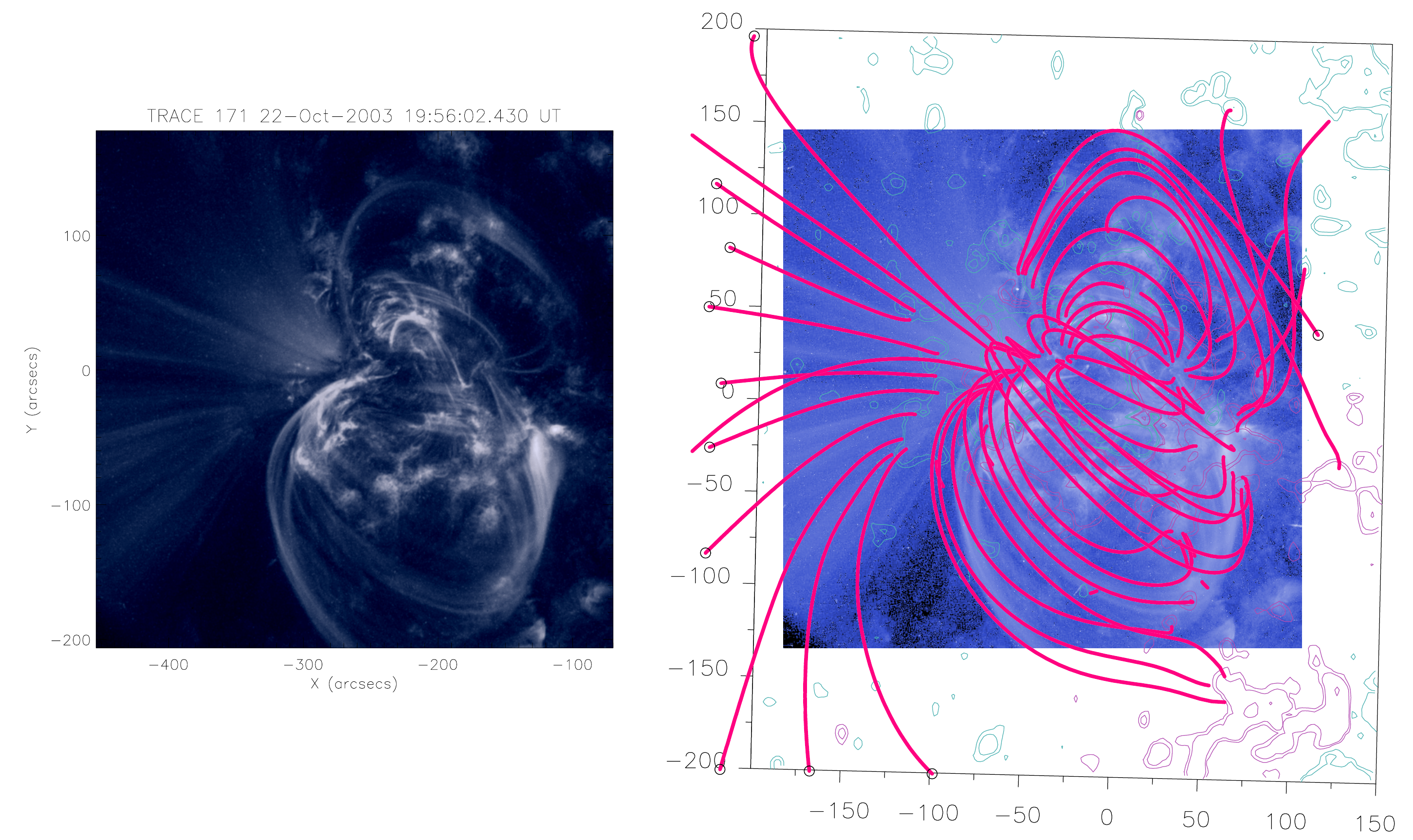}
\caption{Magnetic field model of the AR coronal loops. 
(left panel) TRACE 171 \AA\ image at 19:58 UT on 22 October 2003. (right panel) Magnetic field model of AR 10484 including field lines (pink continuous lines), tracing the EUV loops, overlaid on the same TRACE image. The contours correspond to $\pm$ 50 G, $\pm$ 100 G, and $\pm$ 500 G (magenta for positive and cyan for negative values) and the axes are in Mm. The inclination of the sides of the box (defined in the local frame) indicates that the AR was not at central meridian at that time.}
\label{fig_local}
\end{figure*}

\section{On the origin of flares and jets in AR 10484}
\label{sec_model}

\subsection{The local magnetic-field model}
\label{sec_local}

In this and following sections, we compute the coronal AR magnetic field and compute its topological structure searching 
for clues about the origin of the observed flares and broad jets from 21 to 24 October 2003.
 
The LOS magnetic field of the AR is extrapolated using the linear force free field (LFFF) 
approximation ($\vec{\nabla}\times\vec{B}=\alpha\vec{B}$, where $B$
is the magnetic field and $\alpha$ is the force-free parameter, which is assumed to be constant in the 
entire extrapolation box) with the Fast Fourier Transform (FFT) method described in \citet{Alissandrakis1981}
and \citet{Demoulin1997}.
As a boundary condition for the model, we use the MDI magnetograms closest in time to the observed events. The only free 
parameter  on which the LFFF model depends on, $\alpha$, is chosen so that
the computed field lines match the observed coronal loops. To compare our computed field lines to the observed loops, we select a 171 \AA\ TRACE image. TRACE images were available only in 1600 \AA, 1700 \AA, and white light during the flare and jet on 23 October; therefore, we use the closest in time ($\sim$19:58 UT) 171 \AA\ image on the previous day, which is shown in \fig{local} side by side to the model with computed field lines overlaid on it. The value of $\alpha$ is selected via a recurrent process, as discussed in \citet{Green2002}. The $\alpha$ value that best matches TRACE loops is -1.9\,$\times$\,10$^{-3}$ Mm$^{-1}$.  Our model includes also a transformation of coordinates from the local AR frame to the observed one \citep[see ][]{Demoulin1997}, so that the computed field lines can be directly compared to the observed loops. For more details about how the computational box is selected and its size controlled, see 
\citet{Mandrini2015}.

\subsection{Magnetic topology at the flare and jet location}
\label{sec_BPs}

\subsubsection{BPs: Their general characteristics and relevance}
\label{sec_BP-general}

Typical coronal conditions imply that the magnetic field is force free and frozen into the plasma almost everywhere in solar ARs. Separatrices are an exception where current layers may be formed and energy release is prone to occur. 
Apart from cases in which magnetic null-points are present, separatrices can only appear in a magnetic volume when some field lines touch tangentially the boundary (\ie~the photosphere). This may happen along portions of the photospheric inversion line (PIL) of the magnetic field component normal to this boundary. These portions are called ``bald patches'' or BPs \citep{Titov1993}. At a BP the field lines are curved upwards and the horizontal component of the magnetic field crosses the inversion line from the negative to the positive polarity (\ie~in the opposite way when compared to normal portions of the PIL). The conditions for the existence of a BP were first given by \citet{Seehafer1986} and in more detail by \citet{Titov93}. BPs are present when, at $z=0$ (\ie~the boundary or photosphere), $\vec{B_z}=0$ and $\vec{B} . \vec{\nabla}\vec{B_z}>0$.

BPs define separatrices where current layers can develop \citep[see \eg,][]{Low1988,Vekstein1991,Aly1997}. \citet{Karpen1991} discussed this aspect and showed that the thickness of the low chromosphere could prevent thin current-sheets to form in a BP configuration. However, \citet{Billinghurst1993} explained that strong currents could develop at least near the footpoints of the BP separatrix because a strong concentration of flux tubes could be present in these regions. More recently, \citet{Pariat2009b} showed that currents can accumulate at BPs and their separatrices in MHD numercial simulations; while other simulations 
like those by \citet{Archontis2009}, \citet{Archontis2013}, \citet{Cheung2010}, and \citet{Takasao2015} 
have shown the development of magnetic reconnection at BPs during the build up of ARs. 

When going to observations, there is evidence of the presence of BPs in flares. The first examples were implicitly shown by 
\citet{Seehafer1980} and \citet{Seehafer1985}. \citet{Aulanier1998} found a close correspondence between BP separatrices
and H$\alpha$ and X-ray kernels in a small flare. \citet{Delannee1999} studied a flare associated to 
BPs, where reconnection could have originated a CME. Concerning other type of events, \citet{Fletcher2001} found that transition region brightenings associated with BPs had abundances close to photospheric values. \citet{Mandrini2002} discussed a non-typical scenario in which interacting BPs were related to the formation of arch-filament systems and a H$\alpha$ surge. \citet{Wang2002} found soft X-ray elongated features associated to BPs preceding an X-class flare and CME. \citet{Pariat2004} discussed the relevance of BPs for the emergence of undulatory flux tubes and Ellerman bombs \citep[see also][]{Pariat2009b}, while more recently, a scenario involving U-loops evolving into BPs with reconnection at their location was proposed by \citet{Peter2014} to explain Interface Region Imaging
Spectrograph (IRIS) observations. BPs were also identified in the neighborhood of a jet by \citet{Schmieder2013} close to a magnetic coronal null point. In this particular example, the role of BPs in the jet expulsion was unclear.

\subsubsection{BPs in AR 10484}
\label{sec_BPs-AR}

\begin{figure}[h]
\centering
\includegraphics [width=7.5cm]{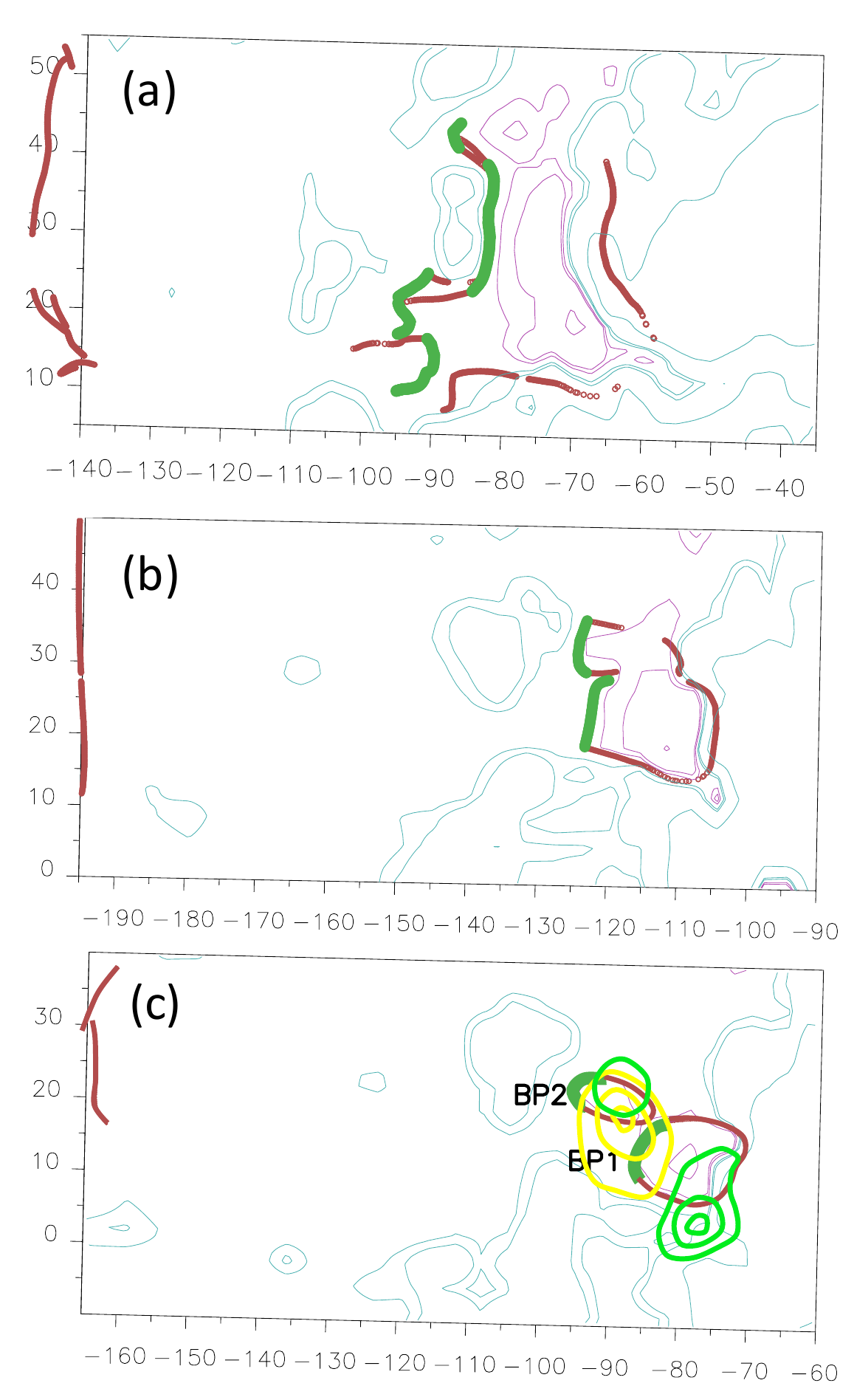}
\caption{Magnetic field models showing the location of the BPs and the intersections with the photosphere of their associated separatrices from 21 to 23 October 2003. (a) Corresponds the MDI map at 20:47 UT on 21 October, (b) to the MDI map at 14:23 UT on 22 October, and (c) to that at 01:35 UT on 23 October. The later is the closest in time to the events we analyze in detail in this article.  In panel (c) we have labeled the BPs as BP1 and BP2 and overlaid the RHESSI contours shown in \fig{trace_rhessi}d (light green contours correspond to the 50--100 keV range and yellow ones to the 10--15 keV range). The three panels are shown from the observer's point of view. The inclination of the sides of the boxes (defined in the local frame) indicates that the AR was not at CM. BPs are shown as the thickest green continuous lines, while the magenta thick continuous lines correspond to the photospheric trace of the BP separatrices. 
The contours correspond to $\pm$ 50 G, $\pm$ 100 G, and $\pm$ 500 G (magenta (cyan) for positive (negative) field values) and the axes are in Mm.}
\label{fig_bp-location}
\end{figure}

As mentioned in \sect{introduction}, numerical models explaining the origin of jets propose as a basis a magnetic configuration above which a magnetic null point and its fan separatix are present \citep[see \eg,][and the review by Raouafi et al. 2016]{Pariat2009a,Pariat2015,Rachmeler2010}. Then,
we first look for the presence of coronal magnetic null points at the location of the flare and jet on 23 October. We find no magnetic null point at that place or in other zone of the AR that could be associated with the flare and broad jet. However, we find a set of two BPs that we call BP1 and BP2. These can be seen in \fig{bp-location}c  together with RHESSI high-energy HXR contours. Furthermore, as discussed in \sect{recurrent-jets}, AR 10484 produced broad jets that erupted in narrow or jet-like CMEs from 21 to 24 October. The recurrency of very similar events at approximately the same location in the AR (see \fig{mdi_evol}) leads us to check if the same magnetic field topology can be found associated with this series of events. The different panels in \fig{bp-location} show the presence of BPs  and the intersection with the photosphere of their associated separatrices at approximately the locations of the base of the jets and brightenings or flares on 21, 22, and 23 October. 

\begin{figure}
\centering
\includegraphics [width=7.5cm]{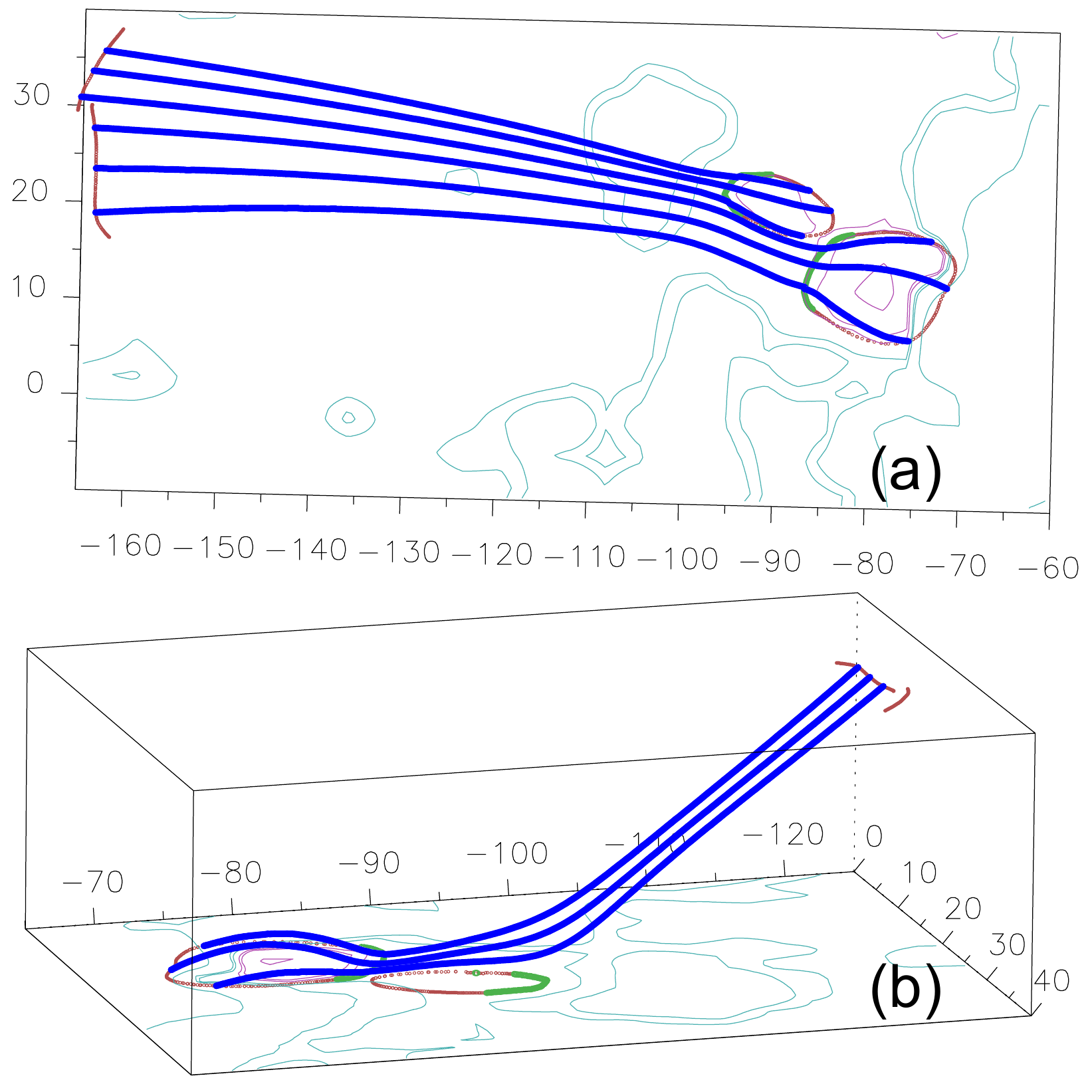}
\caption{Magnetic field model showing field lines that trace the BP separatrices. (a) Corresponds to the observer's point of view. Field lines extending to the east are large-scale and open in our model (ie~ they reach the borders of our computational domain). 
(b) Shows the field lines from a different point of view (rotated by around 180\degree and seen sideways) to clearly illustrate that the separatrix traces to the east are not at the photosphere but reach the top surface of the drawn box. The height of the box is 50 Mm in this figure. For clarity we are only including field lines tracing BP1 separatrix in this panel, a similar field-line tracing is found for BP2. The conventions for the field isocontours, axes, BPs, and separatrix traces are the same as in \fig{bp-location}.}
\label{fig_BP1-lines}
\end{figure}

The magnetic field model, which is the base for the BP computations, is done on each day using as boundary condition the corresponding MDI map and taking the same value of $\alpha$, determined in \sect{local}. The extension and the shape of the BPs varies as the magnetic field evolved, mainly as the positive field polarities were cancelled. 
The BPs appear as separated in different portions. However, we infer that an elongated BP is present on 21 October, which finally evolves into the two portions we call BP1 and BP2 on 23 October. These two BPs are not defining a common separator as described analytically by \citet{Bungey1996} and found in an observed example by \citet{Mandrini2002}.  
In \fig{bp-location} the BPs are shown as the thickest green continuous lines at the photospheric level, while the thick magenta continuous lines correspond to the trace of the BP separatrices. The ones mainly to the west of the positive polarities and within the drawn box (see \fig{bp-location}) lie on the photospheric plane. Those to the east and at the border or out of the drawn box are shown projected on the photosphere.  As an example, \fig{BP1-lines}a shows a set of magnetic field lines in blue color that trace the separatrices of BP1 and BP2. Those reaching the photosphere to the west are low lying and closed, while the ones extending to the east are large-scale and open (as discussed in \sect{PFSS}), they reach the separatices that appear projected on the photosphere (see also \fig{bp-location}c). These separatrix traces lie at the top of our computational domain (see \fig{BP1-lines}b).

 It can be seen  when comparing \fig{bp-location}c with the H$\alpha$ (\fig{halpha}) and TRACE and RHESSI observations (\fig{trace_rhessi}) that the flares kernels and RHESSI high-energy contours are located, within coalignment errors \citep[see][for a discussion on coaligning RHESSI with other satellite instruments]{Berlicki2004} and considering RHESSI's spatial resolution, on the BP separatrices lying at the photospheric level. On the other hand, the low-energy RHESSI contours lie on BP1 and partially on BP2. Furthermore, the magnetic field lines extending to the east in \fig{BP1-lines}a agree well with the direction of the broad jet. This suggests that magnetic reconnection occurring at the BP separatrices could be at the origin of the flare and jet in AR 10484. 

\begin{figure}[h]
\centering
\includegraphics [width=7.5cm]{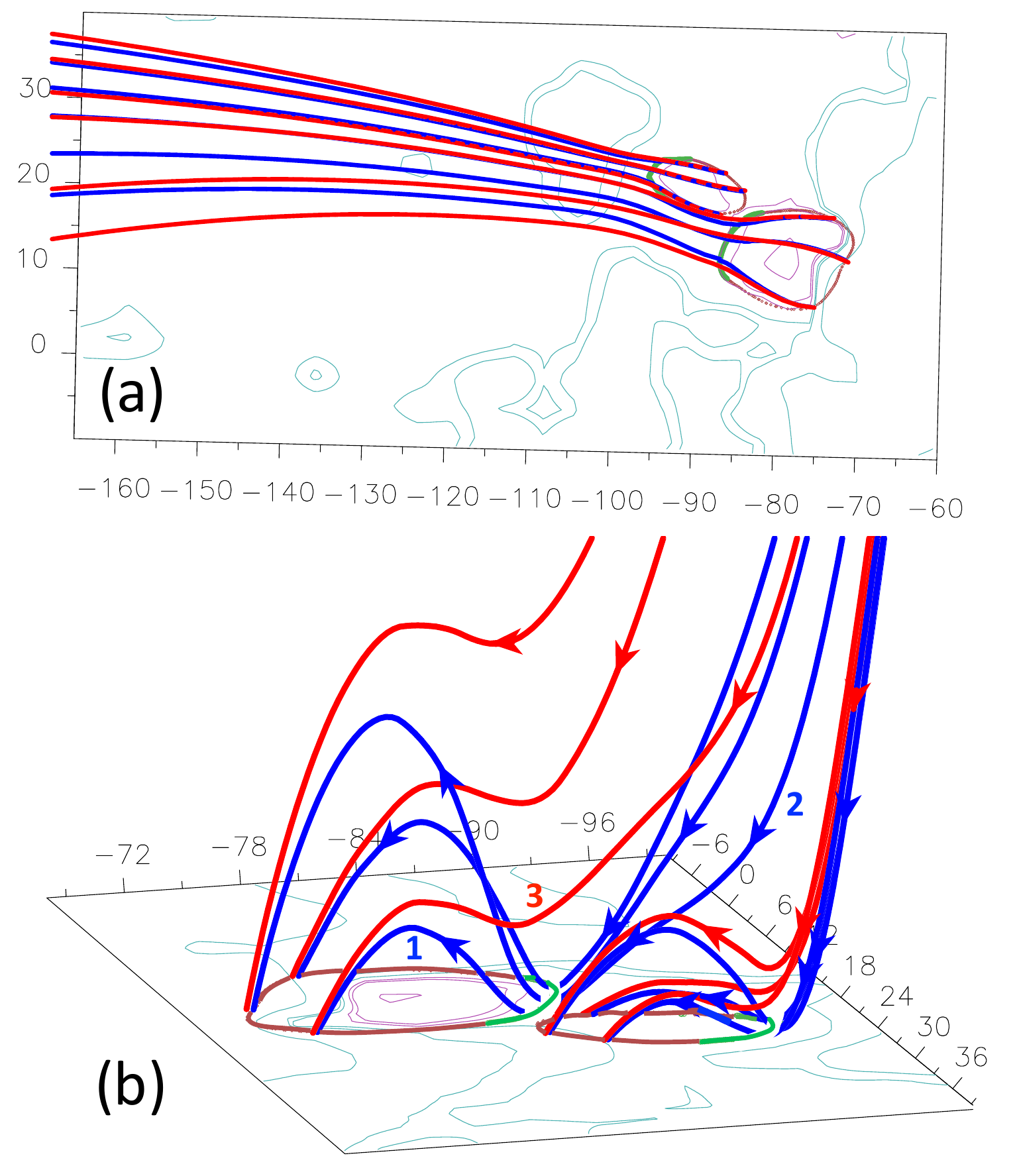}
\vspace*{0.cm}
\caption{Magnetic field model showing the two BPs and a set of pre-reconnected and reconnected field lines as inferred when comparing with observations. 
(a) The blue solid lines would correspond to the situation before reconnection, while the set of red continuous lines issued from the separatrix located at the photospheric level would correspond to the lines after reconnection. This panel is shown in the observer's point of view. (b) Same set of field lines as in panel (a) drawn from a different perspective so that the field connectivity becomes clear. We have also added arrows to the computed lines to indicate the direction of the magnetic field and numbers to some of them to explain how reconnection can proceed (see \sect{BPs-AR}). The height of these field lines has been multiplied by a factor of four so that they can be clearly distinguished. The conventions for the field isocontours, axes, BPs, and separatrices are the same as in \fig{bp-location}}.
\label{fig_bp-lines-recon}
\end{figure}

 \fig{bp-lines-recon} represents a static view directly derived from our coronal field model that shows in a snapshot both pre-reconnected and reconnected field lines. We infer the direction in which reconnection proceeds from the observed event evolution \citep[see e.g. ][and references therein]{Mandrini2015}. In \fig{bp-lines-recon}a we draw in blue and red color a set of pre-reconnected and reconnected field lines, respectively. This panel is shown in the observer's point of view, while in \fig{bp-lines-recon}b the same set of field lines is shown in a perspective that allows to distinguish them clearly. In this figure the closed blue field lines are anchored at the photospheric level by line-tying, while the open ones are anchored only at one end  (at the other end they extend out of our computational domain). The closed blue lines have been computed starting integration very close to the corresponding BP on its western side, while the open ones are computed from the BP eastern side. It is via magnetic reconnection that the blue field lines can evolve into the red field lines with very close footpoints but located on the other side of the photospheric trace of each BP separatrix; the blue field lines are line tied at the photosphere by the gravity force acting on the dense plasma at their bases. Several numerical models (see references in \sect{BP-general}) show that magnetic reconnection can occur in configurations similar to the one we describe (see in particular the sketch in Fig. 8  of Cheung et al. (2010) summarizing the evolution found in their simulation).  

 Our data analysis, combined with modeling and magnetic field topology computation, let us propose two possible scenarios to explain this particular event. These scenarios share the same topology, but the way in which the event is triggered is different.   

A first possibility is that the sustained twisting/shearing motions and flux cancellation can contribute to the continuous formation of a flux rope below the BP closed separatrix. The photospheric evolution can accumulate energy in the constantly formed flux rope and contribute to its destabilization and eruption \citep{vanBallegooijen1989,Aulanier2010}. As the flux rope rises it can force reconnection between oppositely directed field lines above the BPs leading to the formation of a current layer and fast energy release. To illustrate this process we have numbered two pre-reconnected field lines as 1 and 2 in \fig{bp-lines-recon}(b) and the resulting reconnected line as 3.  The result of this process is the impulsive M2.4 flare with kernels located in the close vicinity of the BP separatrices. Plasma in the erupting rope could produce the observed blowout jet as it accesses the open reconnected field lines. This evolution is somehow similar to what happens in numerical simulations of jets having a null point topology, as the ones discussed by \citet{Pariat2009b,Pariat2015}. In these simulations, the photospheric line-tied forcing does not directly stress the fan surface but allows the accumulation of energy in the configuration below the fan, as we propose it happens below the BP separatrix.  

 In the second scenario, we speculate that the continuous twisting/shearing motions and flux cancellation can build up current layers at the BP separatices in a quasistatic way, until impulsive reconnection occurs through a resistive instability
\citep{Billinghurst1993,Pariat2009b}. The process can lead to fast energy release.  As a result, we would observe the impulsive M2.4 flare and, if the reconnection region reaches the open field, the plasma in the open reconnected field lines giving the observed broad jet.
After the event, the configuration relaxes and goes back to a low level energy state. 
However, recent 3D MHD simulations by \citet{Baumann2013}, in which the initial topology contains a null point and the field is stressed by photospheric shearing motions, show that the currents accumulated mainly at the fan separatrix are not enough to give a flare-like energy release as the one needed to explain the observed M2.4 flare. These authors propose that to produce a flare additional free energy should be contained in the initial non-potential state.
Therefore, by analogy with the configuration in \citet{Baumann2013}, in our BP topology we consider that though this second scenario could explain some minor brightenings and collimated jets, it would not be suitable to explain the energetic flare and blowout jet we observe.     
 
\begin{figure}
\centering
\includegraphics[width=7.5cm]{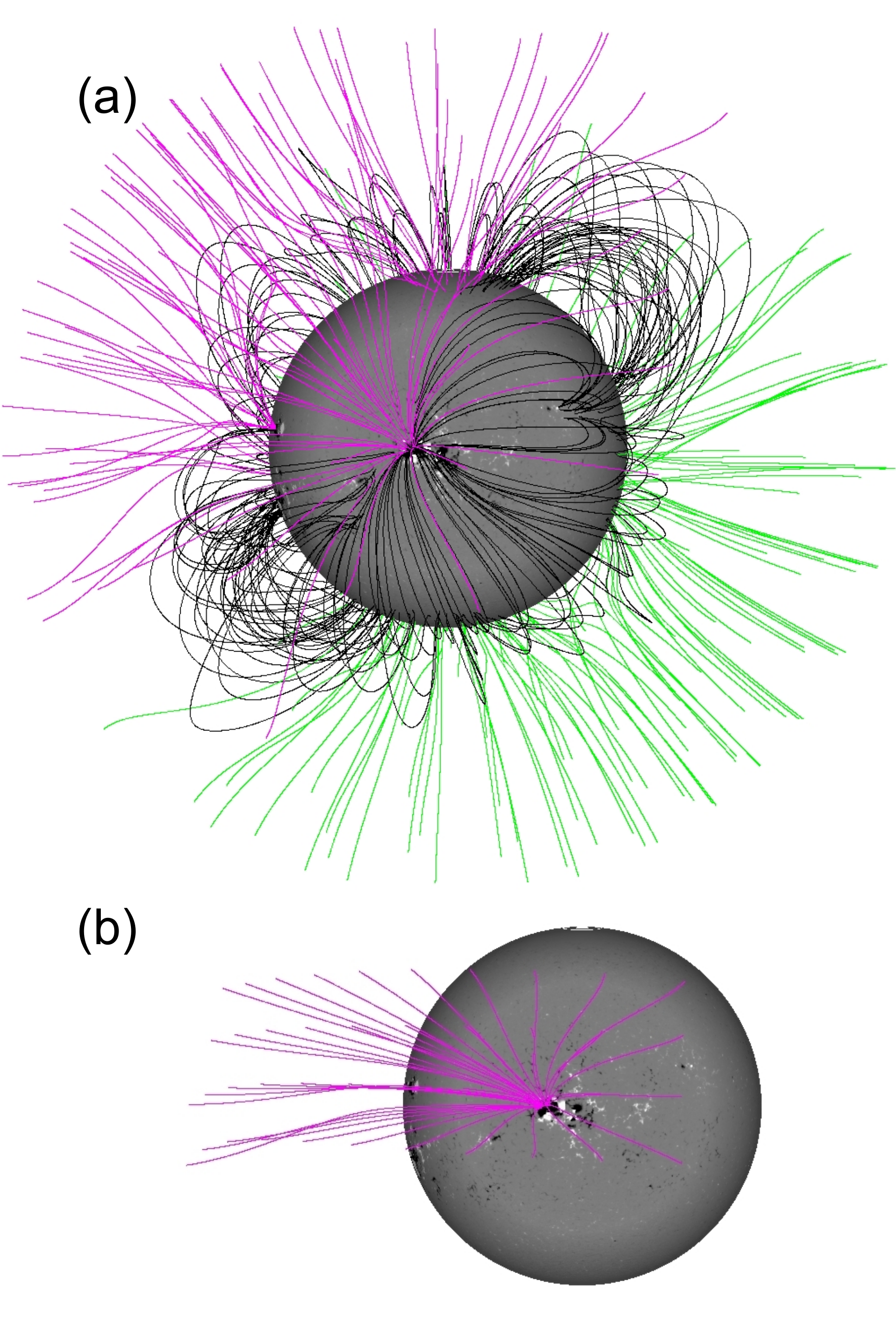}
\caption{ PFSS model of CR 2009 with AR 10484 at CMP (Carrington longitude
$5.5\degree$). (a) Global coronal field. (b) Selection of field lines chosen in a box of $20 \times 20\degree$ with the AR in its center. The field-line colour convention is such that black indicates closed lines and pink (green) corresponds to open lines anchored in the negative-polarity (positive-polarity) field.}
\label{fig_pfss}
\end{figure}

\subsection{The global magnetic-field model}
\label{sec_PFSS}

Our local magnetic field model of AR 10484 shows that field lines associated to the BPs extend to the east and end at the 
top of our computation box. In order to examine what is the behavior of those field lines in the large scale, 
we use a global Potential Field Source Surface (PFSS) model. The model is computed with the Finite Difference 
Iterative Potential-field Solver (FDIPS) code developed by \citet{Toth2011}, which provides more accurate 
results at high-latitude regions than those typically obtained with spherical harmonic expansion methods 
\citep[see][for details on the model]{Mandrini2014}. 
The source surface is set at 2.5 R$_{\rm Sun}$ and the grid is spherical with 150 bins in the radial direction,
180 latitudinal bins, and 360 longitudinal bins. The photospheric boundary condition is prescribed by the 
MDI (polar field corrected) synoptic magnetogram for Carrington rotation (CR) 2009.  

The coronal magnetic field is well described by PFSS models at global scales, except in
regions where high currents are present, such as could be the case around filaments. 
However, for our purpose and considering its limitations, such a model provides a reasonable
description of the connectivity of the field around AR 10484. This is illustrated in \fig{pfss}, 
which shows the result of the model when the AR was at the central meridian. The displayed field lines have been traced using a set of coronal starting points separated by $10\degree$ (in both latitude and longitude) and are located at a height of $\approx 350$ Mm above the photosphere. The selection provides a fairly uniform sampling of the large scale structure around and above the AR. The magnetic field line passing each selected starting point is traced in both directions until the photosphere (1.0 R$_{Sun}$) or the source surface (2.5 R$_{\rm Sun}$) is reached. 
\fig{pfss}a shows the global field model, while in \fig{pfss}b we have selected a set of field lines 
anchored at AR 10484.  From these figures it is clear that 
there is a set of open field lines that could serve as a channel for the plasma ejected in the broad jet that we associate to the narrow eastern CME seen later and depicted in \fig{eit_jet2}.

\section{Summary and conclusions}
\label{sec_conclusions}

A series of recurrent jets were observed in the EUV with SOHO/EIT, with a few seen in the UV with TRACE and in H$\alpha$ at ARIES in Nainital, India, from 21 to 24 October 2003. The jet eruptions originated from similar locations in AR 10484. This site was to the east of a major negative spot around which a smaller negative spot
rotated in the counter-clockwise direction (see \fig{mdi_evol} in \sect{mag} and references therein) forcing the cancellation of small positive polarities located between the rotating spots and to their east. The sustained rotation lasted for at least 48 hs. As a result, the magnetic field at that location was constantly twisted and sheared while negative magnetic helicity was injected.   

The events started with brightenings of different intensities at their bases that were closely followed by the jet eruptions towards the east. The observed jets are broader than standard collimated jets and have different lenghts, with a maximum projected length of $\approx$ 220 Mm (see Table~\ref{recurrent}). For most of the jet events listed in Table~\ref{recurrent}, we identify a coronal counterpart in LASCO images. Only in three cases the association between the jets and the LASCO CMEs is quantitatively established; those are the two events studied in \citet{Yan2012} and the one discussed in \sect{jet-lasco}. The white-light events were all either narrow CMEs or jet-like CMEs. As discussed in \citet{Vourlidas2013}, these type of CMEs may contain filamentary material. Even though we cannot identify a filament or similar cold feature at the base of the jets before the events, we consider that they belong to the blowout class \citep{Moore2010} when taking into account all their observed characteristics. Furthermore, when looking at Table~\ref{recurrent} and comparing with the magnetic field evolution in \fig{mdi_evol}, there seems to be more jets when flux cancellation is more efficient than when the positive flux has almost disappeared. This means that both magnetic field twisting and shearing, with the consequent energy and helicity input into the configuration, plus cancellation play a key role in the observed events.      

As mentioned above, all the blowout jets in Table~\ref{recurrent} were accompanied by brightenings at their bases. In three cases, the energy  built up in the configuration by twisting and shearing of the field is enough to produce three impulsive M-class flares. The location and morphology of the main flare brightenings is similar for the three M flares. Two of these flares, an M1.4 and an M1.2, occurred on 22 October with a difference of one hour and were studied by \citet{Yan2012}. The third flare, the most energetic of the three, is the one analyzed in this article (see \sect{multi}). The M2.4 flare on 23 October started around 10 hours after the second event on 22 October. This flare is characterized by the presence of HXR footpoints seen by RHESSI with a SXR source in between. RHESSI kernels are cospatial with UV and H$\alpha$ kernels and lie at the base of the jet.   

To understand the origin of the homologous blowout jets, small brightenings and flare kernels at their bases, we model the coronal magnetic field of AR 10484 and compute its topology. Our models use MDI magnetograms as boundary conditions during the days when the events were observed. We find no magnetic null points either at the location of the events or at other places in the AR that can be connected to the jets and brightenings. However, we find BPs in our magnetic field models. The location, elongation, and shape of the BPs on the photosphere vary as the magnetic field evolves (see \fig{bp-location}), but they are always at the place from where the jets erupted. 
Magnetic field lines calculated starting integration at the western side of the BPs end at the separatrices on the photosphere, while those starting from the BP eastern sides are very long lines that extend up to the upper surfaces of our computation boxes an end at the separatrices located there (see \figs{BP1-lines}{bp-lines-recon} for the model on 23 October). These later field lines, that open up towards the east limb, are the ones through which the plasma from the jets gets channeled, thus allowing for the observation of the white-light counterparts at eastern equatorial PAs. This is confirmed by a PFSS model of CR 2009 that shows that field lines anchored at the eastern side of AR 10484, at approximately the jet bases, reach the source surface at 2.5 R$_{Sun}$ (see \fig{pfss}), \ie~they are open field lines.      

Combining data analysis with modeling and magnetic field topology computation, we have proposed two possible scenarios to explain the particular event on 23 October at 02:41 UT (see \sect{BPs-AR}). These can be also applied to the blowout jets seen from 21 to 24 October 2003 and the brightenings of varying intensities at their bases (see \sect{recurrent-jets}), which share the same magnetic topology. In the first scenario, twisting/shearing motions and flux cancellation can contribute to the continuous formation of a flux rope below the BP closed separatrix which can eventually erupt and force reconnection between oppositely directed field lines above the BPs. In this particular case, the energy released in the different events depends on the amount stored mainly in the flux rope. Events will occur for as long as both, shearing/twisting motions and flux cancellation, continue. Observations of coronal jets driven by the eruptions of small-scale filaments support this first proposed scenario \citep{Raouafi2012,Kayshap2013,Sterling2015}. In the second scenario, we speculate that current layers can form at the BP separatices, by the continuous twisting/shearing motions and flux cancellation, in a quasistatic way until impulsive reconnection occurs through a resistive instability. The configuration relaxes to a lower energy state after each event. This process can be repeated for as long as the forcing continues. The energy released in the different events would depend on the amount stored between event and event. 
However, the results of 3D MHD simulations by \citet{Baumann2013} show that the currents accumulated (mainly at the fan separatrix in their numerical models) are not enough to produce a flare-like energy release. Therefore, though this scenario could explain the minor brightenings observed from 21 to 24 October, it would not be suitable to explain the blowout jets and energetic flares on 22 and 23 October in AR10484. From this perspective, the first scenario seems the most plausible one.  
 
To compare ejective aspects during two solar rotations of two distinct solar minima, \citet{Cremades2011} used a broad criterion to consider all kinds of white-light coronal ejecta, from thin collimated jets to bright and wide CMEs, going through narrow and faint ejecta. The results discussed in this article support the idea that a continuum of events may exist between standard collimated jets and CMEs. Furthermore, our study is the first clear confirmation that magnetic topologies containing BPs can also produce jets leading the ejected plasma into the open field. In our view, these results should motivate the development of jet models including broader topologies than the classical one based on the presence of magnetic null points.     

\begin{acknowledgements}
CHM, HC, GDC, and FN acknowledge financial support from grants PICT 2012-0973 (ANPCyT) and PIP 2012-01-403 (CONICET).
CHM and GDC recognize grant UBACyT 20020130100321.
HC appreciates support from project UTI4035TC (UTN). AKS  acknowledges the support from RESPOND-ISRO (DOS/PAOGIA2015-16/130/602) project. This work was initiated during a one-month stay of RC and CHM at Paris Observatory, Meudon, for which they acknowledge financial support. We thank Dr. Pascal D\'emoulin for his very helpful comments. 
CHM, HC, and GDC are members of the Carrera del Investigador Cient\'i fico (CONICET). FN is a fellow of CONICET.  
\end{acknowledgements}

\bibliographystyle{aa}
\bibliography{chandra_jets}
\IfFileExists{\jobname.bbl}{}
{\typeout{}
\typeout{****************************************************}
\typeout{****************************************************}
\typeout{** Please run "bibtex \jobname" to optain}
\typeout{** the bibliography and then re-run LaTeX}
\typeout{** twice to fix the references!}
\typeout{****************************************************}
\typeout{****************************************************}
\typeout{}
}

\end{document}